\documentclass[11pt]{article}
\usepackage[pdftex]{graphicx}
\usepackage{graphics}
\usepackage[top=1cm,bottom=2cm,right=2cm,left=2cm]{geometry}
\begin{document}
 \vspace{15pt}

\begin{center}{\Large \bf Completion of the integrable coupling systems}
\end{center}
\begin{center}
{\it Yuqin Yao$^{\dag}$, Chunxia Li$^{\ddag}$ and Shenfeng Shen$^{\S}$\footnote{mathssf@zjut.edu.cn} }
\end{center}
\begin{center}{\small \it$^{\dag}$Department of
  Applied Mathematics, China Agricultural University, Beijing, 100083, PR China\\
 $^{\ddag}$ School of  Mathematical and Sciences, Capital Normal University, Beijing, 100048, PR China \\
$^{\S}$ Department of Applied Mathematics, Zhejiang University of Technology, Hangzhou, 310023, PR China }

\end{center}

\vskip 12pt { \small\noindent\bf Abstract}
 {In this paper, we proposed an procedure to construct the completion of the integrable system by adding a perturbation to the generalized matrix
 problem, which can be used to continuous integrable couplings, discrete integrable couplings and super integrable couplings. As example, we construct
 the completion of the
Kaup-Newell (KN) integrable coupling,   the Wadati-Konno-Ichikawa (WKI) integrable couplingsis, vector  Ablowitz-Kaup-Newell-Segur (vAKNS) integrable couplings, the Volterra integrable couplings, Dirac type integrable couplings and NLS-mKdV type integrable couplings.}

\section{Introduction}


In 1968, Lax proposed the Lax pair approach for studying the KdV equation\cite{a2}. A Lax pair formulation is generally equivalent to a zero curvature equation
\cite{a3}. An  system of the partial differential equations (PDEs) or differential-difference equations (DDEs) is said to be integrable, if it is generated from a continuous zero curvature equation \cite{zce}
\begin{equation}
\label{eqns:04}
U_t-V_x+[U,V]=0,
\end{equation}
or discrete zero curvature equation
\begin{equation}
\label{eqns:04}
U_t+UV-(EV)U=0.
\end{equation}
where the squares matrices,  $U$  and $V$,  are called a Lax pair, often comes from a matrix loop algebra\cite{a4}.
How to seek for new integrable system is an important and interesting topic in the study of mathematical physics.  Integrable couplings, which are  the integrable coupled systems containing given system as their sub-systems, have much richer mathematical structures, such as Hamiltonian structure, infinitely
many symmetries and conservation laws of triangular form \cite{c1}-\cite{c11}. So integrable couplings becomes one of the important topics in the field of integrable systems. At present, based on the zero  curvature presentation, a few approaches have been proposed for constructing integrable couplings, including perturbation, enlarging spectral problem and creating semi-direct sum of the Lie algebra.
Mathematically, a continuous or discrete integrable system reads
\begin{equation}
\label{is}
u_{t}=K(u)=K(u,u_x,\cdots)(or K(u, Eu, E^{-1}u))=J\frac{\delta \mathcal{ H}_1}{\delta u},
\end{equation}
the integrable couplings of  integrable system  (\ref{is})  is a triangular integrable system of the following form \cite{c1}
\begin{equation}
\label{eqns:r}
\left\{
       \begin{array}{ll}
       u_t=K(u), \\
       v_t=S(u,v).
       \end{array}
     \right.
\end{equation}
An example of the integrable couplings is the first-order pertubation\cite{c1}
\begin{equation}
\label{eqns:r}
\left\{
       \begin{array}{ll}
       u_t=K(u), \\
       v_t=K'[v],
       \end{array}
     \right.
\end{equation}
where $K'$ is the Gateaux derivative defined by $K'(u)[v]=\frac{\partial}{\partial\varepsilon}K(u+\varepsilon v,u_x+\varepsilon v_x,\cdots).$

Non-semisimple Lie algebra can decomposed into the semi-direct sum of the semisimple Lie algebra and solvable,  so semi-direct sum of Lie algebra
lay a foundation for constructing integrable couplings. Now we consider the  non-semisimple Lie algebra $\overline{g}$ consisting of the
square matrices $M(U,U_1)$ in the block form, i. e.
\begin{equation}
\label{eqns:m}
M(U,U_1)=\left[
      \begin{array}{cc}
        U & U_1\\
      0 & U \\
      \end{array}
    \right],
\end{equation}
where $U$ and $U_1$ are square matrices of the same order. This Lie algebra $\overline{g}$ possess the  semi-direct sum decomposition $\overline{g}=g \oplus g_c$, where $g=M(U,0)$ is semisimple and $g_c=M(0,U_1)$ is solvable. The notation of semi-direct sum implies that the subalgebra $g$ and $g_c$ satisfy $[g,g_c]\subseteq g_c$, with  $[.,.]$ denoting Lie bracket of $\overline{g}$.  Also the closure property between $g$ and $g_c$ under the matrix multiplication is  required, that is to say $gg_c,g_cg\subseteq g_c$.

In our paper, based on the  non-semisimple Lie algebra $\overline{g}$, we propose a method to construct generalized integrable coupled system
\begin{equation}
\label{eqns:gc}
\left\{
       \begin{array}{ll}
       u_t=\overline{K}(u,v), \\
       v_t=\overline{S}(u,v),
       \end{array}
     \right.
\end{equation}
which can be reduced to the standard integrable couplings  (\ref{eqns:r}). In this sense, we call the system   (\ref{eqns:gc})   "completion of the integrable couplings ". In this paper, we will apply this method to  continuous integrable couplings, discrete integrable couplings and super integrable couplings.Concretely, we construct
 the completion of the
Kaup-Newell (KN) integrable coupling,   the Wadati-Konno-Ichikawa (WKI) integrable couplingsis, vector  Ablowitz-Kaup-Newell-Segur (vAKNS) integrable couplings, the Volterra integrable couplings, Dirac type integrable couplings and NLS-mKdV type integrable couplings. What's more, we take the complete integrable system of the KN integrable coupling as an example to show that its Hamiltonian structure can be constructed by variable identity.

\section{Completion of the continuous integrable couplings}
\subsection{Completion of the KN integrable couplings}
\subsubsection{Complete integrable system of the KN integrable couplings}
Consider the generalized spectral problem
\begin{equation}
\label{eqns:0}
\phi_x=\overline{U}\phi,~u=\left[
                  \begin{array}{c}
                    p \\
                    q \\
                    r\\
                    s
                  \end{array}
                \right],~\phi=\left[
                                \begin{array}{c}
                                  \phi_1 \\
                                  \phi_2 \\
                                  \phi_3\\
                                  \phi_4
                                \end{array}
                              \right],
\end{equation}
where the spectral matrix  $\overline{U}$ is chosen as
\begin{equation}
\label{eqns:K2} \overline{U}=\left[
      \begin{array}{cc}
        U &U_1\\
      0 &U \\
      \end{array}
    \right]=\left[
      \begin{array}{cc|cc}
       \lambda^2+h &\lambda p &0& \lambda r\\
      \lambda q & -\lambda^2-h & \lambda s &0 \\ \hline
      0&0& \lambda^2+h &\lambda\\
      0&0& \lambda q & -\lambda^2-h
      \end{array}
    \right],~~~h=\varepsilon(ps+qr)
\end{equation}
 here a nonlinear perturbation term $h$ is added to the spectral matrix. When $\varepsilon=0$, it reduces to the case of KN integrable couplings\cite{ckn1,ckn2}. From the above spectral matrix, we can work out the  complete system of the KN integrable couplings.

Now let us assume that $W$ has the following form
\begin{equation}
\label{eqns:w}
\overline{W}=\left[
      \begin{array}{cc}
        W &W_1\\
      0 &W \\
      \end{array}
    \right]=\left[
      \begin{array}{cc|cc}
       a &b &f&d\\
      c & -a &g &-f \\ \hline
      0&0& a&b\\
      0&0&c & -a
      \end{array}
    \right]
\end{equation}
 and solving the stationary zero curvature equation
$
\overline{W}_x=[\overline{U},\overline{W}]
$
gives\begin{equation}
\label{eqns:dtkn}
\left\{
       \begin{array}{ll}
        a_x=\lambda pc-\lambda qb, \\
        b_x=2\lambda^{2}b+2hb-2\lambda pa,\\
       c_x=-2\lambda^{2}c-2hc+2\lambda qa,\\
       d_x=2\lambda^{2}d-2\lambda ar- 2\lambda fp+2hd,\\
       f_x=-\lambda bs+\lambda rc- \lambda dq+\lambda gp,\\
      g_x=-2\lambda^{2}g+2\lambda as+2\lambda qf-2gh.\\
       \end{array}
     \right.
\end{equation}
 Further, let $a,b,c$ possess the Laurent expansions:
$$a=\sum_{i\geq 0}a_i\lambda^{-2i},~b=\sum_{i\geq 0}b_i\lambda^{-2i-1},~c=\sum_{i\geq 0}c_i\lambda^{-2i-1},
$$$$ f=\sum_{i\geq 0}f_i\lambda^{-2i},~d=\sum_{i\geq 0}d_i\lambda^{-2i-1},~g=\sum_{i\geq 0}g_i\lambda^{-2i-1}.
$$
Equivalently,  the system (\ref{eqns:dtkn}) leads to the recursion relations
\begin{equation}
\label{eqns:dt1kn}
\left\{
       \begin{array}{ll}
        a_{ix}= pc_i- qb_i, \\
        b_{i+1}=\frac{1}{2}b_{ix}+pa_{i+1}-hb_i,\\
       c_{i+1}=-\frac{1}{2}c_{ix}+qa_{i+1}-hc_i,\\
       d_{i+1}=\frac{1}{2}d_{ix}+ra_{i+1}+pf_{i+1}-hd_i,~~~~~i\geq 0\\
       f_{ix}=-s b_i+ rc_i- q d_i+pg_i,\\
      g_{i+1}=-\frac{1}{2}g_{ix}+sa_{i+1}+qf_{i+1}-hg_i.\\
       \end{array}
     \right.
\end{equation}
upon taking the initial data
$$a_0=1,~b_0=p,~c_0=q,~d_0=r,~g_0=s, ~f_0=1.$$
 To guarantee the uniqueness of
 $\{a_i,b_i,c_i,d_i,f_i,g_i|i\geq 1\}$ , we impose the conditions on constants of integration.
 $$a_i\mid_{u=0}=b_i\mid_{u=0}=c_i\mid_{u=0}=d_i\mid_{u=0}=f_i\mid_{u=0}=g_i\mid_{u=0}.$$
 Thus, the first two sets can be listed as follows:
$$a_1=-\frac{pq}{2},~b_1=\frac{1}{2}p_x-\frac{1}{2}p^2q-\varepsilon(ps+qr)p,~c_1=-\frac{1}{2}q_x-\frac{1}{2}pq^2-\varepsilon(ps+qr)q,~f_1=-\frac{1}{2}(ps+qr),
$$$$d_1=\frac{1}{2}r_x-pqr-\frac{1}{2}p^2s-\varepsilon(ps+qr)r,~g_1=-\frac{1}{2}s_x-pqs-\frac{1}{2}q^2r-\varepsilon(ps+qr)s,~~~~~~~~~~~~~~~~~~~~~~~$$
$$a_2=\frac{1}{4}(pq_x-qp_x)+\frac{3}{8} p^{2}q^2+\varepsilon(ps+qr)pq,~~~~~~~~~~~~~~~~~~~~~~~~~~~~~~~~~~~~~~~~~~~~~~~~~~~~~~~~~~~~~~~~~~~~$$
$$b_2=\frac{1}{4}p_{xx}-(\frac{3}{4}qp+\varepsilon(ps+qr))p_x+\frac{3}{8} p^{3}q^{2}+\frac{3}{2}\varepsilon(ps+qr)p^2q+\varepsilon^2(ps+qr)^2p,~~~~~~~~~~~~~~~~~~~~~~~$$
$$c_2=\frac{1}{4}q_{xx}+(\frac{3}{4}qp+\varepsilon(ps+qr))q_x+\frac{3}{8} q^{3}p^{2}+\frac{3}{2}\varepsilon(ps+qr)pq^2+\varepsilon^2(ps+qr)^2q,~~~~~~~~~~~~~~~~~~~~~~~$$
$$f_2=\frac{1}{4}(ps_x-sp_x-qr_x+rq_x)+\frac{3}{4}(sqp^2+prq^2)+\varepsilon(ps+qr)^2,~~~~~~~~~~~~~~~~~~~~~~~~~~~~~~~~~~~~~~~~~$$
$$d_2=\frac{1}{4}r_{xx}-\frac{3}{4}(spp_x+rqp_x+pqr_x)-\varepsilon r_x+\frac{3}{4}sqp^3+\frac{9}{8}rp^2q^2+\varepsilon(ps+qr)(\frac{3}{2}p^2s+3rpq)+\varepsilon^2(ps+qr)^2r$$
$$g_2=\frac{1}{4}s_{xx}+\frac{3}{4}(pqs_x+spq_x+rqq_x)+\varepsilon s_x+\frac{3}{4}rpq^3+\frac{9}{8}sp^2q^2+\varepsilon(ps+qr)(\frac{3}{2}q^2r+3spq)+\varepsilon^2(ps+qr)^2s$$
Noting
\begin{equation}
\label{eqns:v}
V^{[m]}=\lambda(\lambda^{2m+1}W)_++\triangle_m,~m\geq 0,
\end{equation}
where the modification term $\triangle_m=\left[
      \begin{array}{cccc}
       \delta_m &0 &0& 0\\
    0 & - \delta_m & 0 &0 \\
      0&0&  \delta_m&0\\
      0&0& 0& - \delta_m
      \end{array}
    \right],$ $\delta_m$ is undetermined function and $P_+$ denotes the polynomial part of $P$, the corresponding zero curvature equations
\begin{equation}
\label{eqns:zce}
\overline{U}_{t_{m}}-V^{[m]}_{x}+[\overline{U},V^{[m]}]=0,~m\geq0,
\end{equation}
give rise to a hierarchy
\begin{equation}
\label{eqns:hkn}
\left\{
       \begin{array}{ll}
       p_{t}=2(-b_{m+1}+pa_{m+1}+p\delta_{m}), \\
       q_{t}=2(c_{m+1}-qa_{m+1}-q\delta_{m}),~m\geq 0.\\
        r_{t}=2(-d_{m+1}+ra_{m+1}+pf_{m+1}+r\delta_{m}),\\
         s_{t}=2(g_{m+1}-sa_{m+1}-qf_{m+1}-s\delta_{m}),\\
     h_{t}=\delta_{mx}.
       \end{array}
     \right.
\end{equation}
Based on  (\ref{eqns:dt1kn}) and (\ref{eqns:hkn}), we can have
$$\delta_{mx}=h_t=\varepsilon (p_ts+ps_t+q_tr+qr_t)$$
$$=\varepsilon[2s(-b_{m+1}+pa_{m+1}+p\delta_{m})+2p(g_{m+1}-sa_{m+1}-qf_{m+1}-s\delta_{m})$$$$~~~~+2r(c_{m+1}-qa_{m+1}-q\delta_{m})+2q(-d_{m+1}+ra_{m+1}+pf_{m+1}+r\delta_{m})]$$
$$=2\varepsilon (-sb_{m+1}+pg_{m+1}+rc_{m+1}-qd_{m+1})=2\varepsilon f_{m+1,x}.~~~~~~~~~~~~~~~~~~$$
So, we choose that $\delta_{m}=2\varepsilon f_{m+1}$, and generate the complete system of the KN integrable couplings:
\begin{equation}
\label{eqns:cskn}\left[
      \begin{array}{c}
       p_t\\
    q_t \\
     r_t\\
    s_t
      \end{array}
    \right]=\left[
      \begin{array}{c}
       2(-b_{m+1}+pa_{m+1}+2\varepsilon pf_{m+1})\\
    2(c_{m+1}-qa_{m+1}-2 \varepsilon qf_{m+1}) \\
      2(-d_{m+1}+ra_{m+1}+pf_{m+1}+2\varepsilon rf_{m+1}),\\
     2(g_{m+1}-sa_{m+1}-qf_{m+1}-2\varepsilon sf_{m+1})
      \end{array}
    \right],~~~~~m\geq 0.
\end{equation}
A nonlinear equation in the above new system is given
$$p_t=-\frac{1}{2}p_{xx}+(\frac{1}{2}p^2-\varepsilon pr)q_x+(pq-\varepsilon ps+2\varepsilon (ps+qr))p_x-\varepsilon pqr_x+\varepsilon p^2s_x-\varepsilon (ps+qr)p^2q$$$$+\varepsilon (3p^2q^2r+p^3sq)+\varepsilon^2(ps+qr)p^2s-\varepsilon^2 (ps+qr)^2p,~~~~~~~~~~~~~~~~~~~~~~~~~~~~~~~~~$$
$$q_t=\frac{1}{2}q_{xx}+(q^2+\varepsilon qs)p_x+(\frac{1}{2}pq-\varepsilon qr+2\varepsilon (ps+qr))q_x-\varepsilon pqs_x+\varepsilon q^2r_x-3\varepsilon pq(q^2r+pqs)$$$$-4\varepsilon^2 (ps+qr)^2-\frac{3}{4}q^3p^2-\varepsilon (ps+qr)pq^2+2\varepsilon^2 (ps+qr)^2q,~~~~~~~~~~~~~~~~~~~~~~~~~~~~~$$
$$r_t=-\frac{1}{2}r_{xx}+(ps+qr+\varepsilon rs)p_x+(pq+2\varepsilon (2ps+3qr))r_x+(\frac{1}{2}p^2-\varepsilon pr)s_x+(pr-\varepsilon r^2)q_x$$$$-3\varepsilon pqr(qr+ps)-\varepsilon(ps+qr)p(ps+2qr)-6\varepsilon^2 (ps+qr)^2r,~~~~~~~~~~~~~~~~~~~~~~~~~~$$
$$s_t=\frac{1}{2}s_{xx}+(qs+\varepsilon s^2)p_x+(ps+qr-\varepsilon sr)q_x+(\frac{1}{2}q^2+\varepsilon qs)r_x+(pq+\varepsilon(ps+2qr))s_x~~~~$$$$-3\varepsilon pqs(qr+ps)-2\varepsilon^2 (ps+qr)^2s+\varepsilon(ps+qr)q(2ps+qr).~~~~~~~~~~~~~~~~~~~~~~~~~~$$

\subsubsection{Hamiltonian structure of the complete system}
In the section, we will construct the Hamiltonian structure for the complete system (\ref{eqns:cskn}) by using the variational identity\cite{vi1}
\begin{equation}
\label{eqns:it}
\frac{\delta}{\delta u}\int tr(W\frac{\partial U_1}{\partial \lambda}+W_1\frac{\partial U}{\partial \lambda})dx=\lambda^{-\gamma}\frac{\partial}{\partial \lambda} \lambda^{\gamma}tr(W\frac{\partial U_1}{\partial \overline{u}}+W_1\frac{\partial U}{\partial \overline{u}})
\end{equation}

Direct computation gives
$$ tr(W\frac{\partial U_1}{\partial \lambda}+W_1\frac{\partial U}{\partial \lambda})=4f\lambda+sb+rc+qd+pg,~tr(W\frac{\partial U_1}{\partial p}+W_1\frac{\partial U}{\partial p})=2\varepsilon sf+g\lambda,~~~~~~~~~~~~~~~~$$
$$ tr(W\frac{\partial U_1}{\partial q}+W_1\frac{\partial U}{\partial q})=2\varepsilon r f+d\lambda,~tr(W\frac{\partial U_1}{\partial r}+W_1\frac{\partial U}{\partial r})=2\varepsilon qf+c\lambda,~tr(W\frac{\partial U_1}{\partial s}+W_1\frac{\partial U}{\partial s})=2\varepsilon p f+b\lambda.$$

Substitute the above results into the variational identity (\ref{eqns:it}), we have
\begin{equation}
\label{eqns:it1}
\frac{\delta}{\delta \overline{u}}\int (4f\lambda+sb+rc+qd+pg)dx=\lambda^{-\gamma}\frac{\partial}{\partial \lambda} \lambda^{\gamma}\left[
      \begin{array}{c}
       2\varepsilon sf+g\lambda\\
    2\varepsilon r f+d\lambda \\
      2\varepsilon qf+c\lambda,\\
     2\varepsilon p f+b\lambda
      \end{array}
    \right],~
\end{equation}

Balancing coefficients of the $\lambda^{-2m-3}$ gives
\begin{equation}
\label{eqns:it2}
\frac{\delta}{\delta \overline{u}}\int (4f_{m+2}+sb_{m+1}+rc_{m+1}+qd_{m+1}+pg_{m+1})dx=(\gamma-2m-2)\left[
      \begin{array}{c}
       2 \varepsilon sf_{m+1}+g_{m+1}\\
    2\varepsilon r f_{m+1}+d_{m+1}\\
      2 \varepsilon qf_{m+1}+c_{m+1},\\
     2\varepsilon  p f_{m+1}+b_{m+1}
      \end{array}
    \right].~
\end{equation}
Consider the case of $m=0$, we have $\gamma=0$. Thus, we have
\begin{equation}
\label{eqns:it3}
\frac{\delta}{\delta \overline{u}}\int -\frac{4f_{m+2}+sb_{m+1}+rc_{m+1}+qd_{m+1}+pg_{m+1}}{2m+2}dx=\left[
      \begin{array}{c}
       2\varepsilon sf_{m+1}+g_{m+1}\\
    2 \varepsilon r f_{m+1}+d_{m+1}\\
      2\varepsilon qf_{m+1}+c_{m+1},\\
     2 \varepsilon p f_{m+1}+b_{m+1}
      \end{array}
    \right].~
\end{equation}
The Hamiltonian structure can be given
 \begin{equation}\label{h2}
u_{t_m}=K_m=J\frac{\delta \mathcal{H}_m}{\delta u},
\end{equation}
with a Hamiltonian operator
$$J=\left[\begin{array}{cccc}
                                          0 & 0&-2p\partial^{-1}p &2+2p\partial^{-1}q \\
                                       0 & 0&-2+2q\partial^{-1}p &-2q\partial^{-1}q \\
                                       -2p\partial^{-1}p &2+2p\partial^{-1}q & -2r\partial^{-1}p-2p\partial^{-1}r &2r\partial^{-1}q+2p\partial^{-1}s\\
                                       -2+2q\partial^{-1}p & -2q\partial^{-1}q & 2s\partial^{-1}p+2q\partial^{-1}r & -2s\partial^{-1}q-2q\partial^{-1}s
                                        \end{array}
                                      \right],$$
  and the Hamiltonian functions
  $$\mathcal{H}_m=\int -\frac{4f_{m+2}+sb_{m+1}+rc_{m+1}+qd_{m+1}+pg_{m+1}}{2m+2}dx.$$

Using recursion relations (\ref{eqns:dt1kn}), we can obtain the recursion operator $\Phi=(\Phi_{ij})_{4\times 4}$  through
$$\left[
      \begin{array}{c}
       2\varepsilon sf_{m+1}+g_{m+1}\\
    2\varepsilon r f_{m+1}+d_{m+1}\\
      2\varepsilon qf_{m+1}+c_{m+1},\\
     2 \varepsilon p f_{m+1}+b_{m+1}
      \end{array}
    \right]=\left[
                \begin{array}{cccc}
                 \Phi_{11} & \Phi_{12} & \Phi_{13} & \Phi_{14}  \\
                 \Phi_{21} & \Phi_{22} & \Phi_{23} & \Phi_{24}\\
                 \Phi_{31} & \Phi_{32} & \Phi_{33} & \Phi_{34}\\
                 \Phi_{41} & \Phi_{42} & \Phi_{43} & \Phi_{44}
                  b_m \\
                \end{array}
              \right]\left[
      \begin{array}{c}
       2\varepsilon sf_{m}+g_{m}\\
    2\varepsilon r f_{m}+d_{m}\\
      2\varepsilon qf_{m}+c_{m},\\
     2 \varepsilon p f_{m}+b_{m}
      \end{array}
    \right]
$$
with $$\Phi_{11}=-(\varepsilon s\partial^{-1} p+\frac{1}{2}+\frac{1}{2} pq)\partial-(2\varepsilon s\partial^{-1} p+1+ pq)h+\Xi_1 \partial^{-1}p,~~~~~~~~~~~~~~~~~~~~~~~$$
$$\Phi_{12}=-(\varepsilon s\partial^{-1} q+\frac{1}{2} q^2)\partial+(2\varepsilon s\partial^{-1} q+q^2)h-\Xi_1 \partial^{-1}q,~~~~~~~~~~~~~~~~~~~~~~~~~~~~~~~~~~~$$
$$\Phi_{13}=-(\varepsilon s\partial^{-1} r+\frac{1}{2}s\partial^{-1} p+\frac{1}{2} qr)\partial-(2\varepsilon s\partial^{-1} r+s\partial^{-1} p+ qr)h+\Xi_1 \partial^{-1}r,~~~~~~~~~~$$
$$\Phi_{14}=-(\varepsilon s\partial^{-1} s+\frac{1}{2}s\partial^{-1} q+\frac{1}{2} qs)\partial+(2\varepsilon s\partial^{-1} s+s\partial^{-1} q+ qs)h-\Xi_1 \partial^{-1}s,~~~~~~~~~~$$
$$\Phi_{21}=-(\varepsilon r\partial^{-1} p+\frac{1}{2}p\partial^{-1} p)\partial-(2\varepsilon r\partial^{-1} p+p\partial^{-1} p)h+\Xi_2 \partial^{-1}p,~~~~~~~~~~~~~~~~~~~~~~~~~$$
$$\Phi_{22}=-(\varepsilon r\partial^{-1} q+\frac{1}{2}p\partial^{-1} q)\partial+(2\varepsilon r\partial^{-1} q+p\partial^{-1} q)h-\Xi_2 \partial^{-1}q,~~~~~~~~~~~~~~~~~~~~~~~~~$$
$$\Phi_{23}=-(\varepsilon r\partial^{-1} r+\frac{1}{2}q\partial^{-1} p+\frac{1}{2}p\partial^{-1} r)\partial-(2\varepsilon r\partial^{-1} r+q\partial^{-1} p+p\partial^{-1} r)h+\Xi_2 \partial^{-1}r,$$
$$\Phi_{24}=-(\varepsilon r\partial^{-1} s+\frac{1}{2}q\partial^{-1} q+\frac{1}{2}p\partial^{-1} s)\partial+(2\varepsilon r\partial^{-1} s+q\partial^{-1} q+p\partial^{-1} s)h-\Xi_2 \partial^{-1}s,$$
$$\Phi_{31}=-(\varepsilon q\partial^{-1} p\partial+2\varepsilon q\partial^{-1} ph+\Xi_3 \partial^{-1}p,~~~~~~~~~~~~~~~~~~~~~~~~~~~~~~~~~~~~~~~~~~~~~~~~~~~~$$
$$\Phi_{32}=-(\varepsilon q\partial^{-1} q\partial+2\varepsilon q\partial^{-1} qh-\Xi_3 \partial^{-1}q,~~~~~~~~~~~~~~~~~~~~~~~~~~~~~~~~~~~~~~~~~~~~~~~~~~~~~$$
$$\Phi_{33}=-(\varepsilon q\partial^{-1} r+\frac{1}{2}q\partial^{-1} p+\frac{1}{2})\partial-(2\varepsilon q\partial^{-1} r+q\partial^{-1} p+1)h-\Xi_3 \partial^{-1}r,~~~~~~~~~~~~~$$
$$\Phi_{34}=-(\varepsilon q\partial^{-1} s+\frac{1}{2}q\partial^{-1} q)\partial-(2\varepsilon q\partial^{-1} s+q\partial^{-1} q)h-\Xi_3 \partial^{-1}s,~~~~~~~~~~~~~~~~~~~~~~~~$$
$$\Phi_{41}=-(\varepsilon p\partial^{-1} p\partial-2\varepsilon p\partial^{-1} ph+\Xi_4 \partial^{-1}p,~~~~~~~~~~~~~~~~~~~~~~~~~~~~~~~~~~~~~~~~~~~~~~~~~~$$
$$\Phi_{42}=-(\varepsilon p\partial^{-1} q\partial+2\varepsilon p\partial^{-1} qh-\Xi_4 \partial^{-1}q,~~~~~~~~~~~~~~~~~~~~~~~~~~~~~~~~~~~~~~~~~~~~~~~~~~~$$
$$\Phi_{43}=-(\varepsilon p\partial^{-1} r+\frac{1}{2}p\partial^{-1} p)\partial-(2\varepsilon p\partial^{-1} r+p\partial^{-1} p)h+\Xi_4 \partial^{-1}r,~~~~~~~~~~~~~~~~~~~$$
$$\Phi_{44}=-(\varepsilon p\partial^{-1} s+\frac{1}{2}p\partial^{-1} q-\frac{1}{2})\partial+(2\varepsilon p\partial^{-1} s+p\partial^{-1} q-1)h-\Xi_4\partial^{-1}s,~~~~~~~~~$$
where {\small$$\Xi_1=2\varepsilon [(\varepsilon s\partial^{-1} s+\frac{1}{2}qs+\frac{1}{2}s\partial^{-1} q)\partial p+(\varepsilon s\partial^{-1} r+\frac{1}{2}qr+\frac{1}{2}s\partial^{-1} p)\partial q+(\varepsilon s\partial^{-1} q+\frac{1}{2}q^2)\partial r+(\varepsilon s\partial^{-1} p+\frac{1}{2}+\frac{1}{2}pq)\partial s+hs],$$
$$\Xi_2=2\varepsilon [(\varepsilon r\partial^{-1} s+\frac{1}{2}q\partial^{-1} q+\frac{1}{2}p\partial^{-1} s)\partial p+(\varepsilon r\partial^{-1} r+\frac{1}{2}q\partial^{-1} p+\frac{1}{2}p\partial^{-1} r)\partial q+(\varepsilon r\partial^{-1} q+p\partial^{-1} q)\partial r+(\varepsilon r\partial^{-1} p+p\partial^{-1} p)\partial s],$$
$$\Xi_3=2\varepsilon^2 (q\partial^{-1} s\partial p+q\partial^{-1} r\partial q+q\partial^{-1} q\partial r+q\partial^{-1} p\partial s),~~~~~~~~~~~~~~~~~~~`
~~~~~~~~~~~~~~~~~~~~~~~~~~~~~~~~~~~~~~~~~~~~~$$
$$\Xi_4=2\varepsilon [(\varepsilon p\partial^{-1} s+\frac{1}{2}p\partial^{-1} q-\frac{1}{2})\partial p+(\varepsilon p\partial^{-1} r+\frac{1}{2}p\partial^{-1} p)\partial q+\varepsilon p\partial^{-1} q\partial r+\varepsilon p\partial^{-1} p\partial s+ph].~~~~~~~~~~~~~~~~$$}
Thus, we
 show that all members in the hierarchy (\ref{eqns:cs}) are bi-Hamiltonian
\begin{equation}\label{bh}
u_{t_{m}}=K_m=J\frac{\delta \mathcal{H}_m}{\delta u}=M\frac{\delta\mathcal{H}_{m-1}}{\delta u},~m\geq 0,
\end{equation}
where the second Hamiltonian operator is given by
$M=J\Phi.$

 So, the soliton hierarchy (\ref{eqns:cs}) is Liouville integrable. Particularly, it possesses infinitely many commuting conserved functions and symmetries:
$$\{H_k,H_l\}_J=\int(\frac{\delta H_k}{\delta u})^{T}J\frac{\delta H_l}{\delta u}dx=0,~k,l\geq 0,$$
$$\{H_k,H_l\}_M=\int(\frac{\delta H_k}{\delta u})^{T}M\frac{\delta H_l}{\delta u}dx=0,~k,l\geq 0,$$
and
$$[K_k,K_l]=K'_k(u)[K_l]-K'_l(u)[K_k]=0,~k,l\geq 0.$$
These commuting relations are also consequences of the Virasoro algebra of Lax operators\cite{ma7}.

\subsection{Completion of the WKI integrable couplings}
Consider the following generalized spectral problem
\begin{equation}
\label{eqns:0}
\phi_x=\overline{U}\phi,~u=\left[
                  \begin{array}{c}
                    p \\
                    q \\
                    r\\
                    s
                  \end{array}
                \right],~\phi=\left[
                                \begin{array}{c}
                                  \phi_1 \\
                                  \phi_2 \\
                                  \phi_3\\
                                  \phi_4
                                \end{array}
                              \right],
\end{equation}
with the spectral matrix  $\overline{U}$ as follows
\begin{equation}
\label{eqns:K2} \bar{U}=\left[
      \begin{array}{cc}
        U &U_1\\
      0 &U \\
      \end{array}
    \right]=\left[
      \begin{array}{cc|cc}
       \lambda+h &\lambda p &0& \lambda r\\
      \lambda q & -\lambda-h & \lambda s &0 \\ \hline
      0&0& \lambda+h &\lambda p\\
      0&0& \lambda q & -\lambda-h
      \end{array}
    \right],~~~h=\varepsilon(ps+qr)_x.
\end{equation}
 Here, a nonlinear perturbation term $h$ is added to the spectral matrix. When $\varepsilon=0$, it reduces the case of WKI integrable couplings\cite{wki1,wki2}. From  thus spectral matrix, we can work out the  complete system of the WKI integrable couplings by standard procedure.

Solving the stationary zero curvature equation
$
\overline{W}_x=[\overline{U},\overline{W}]
$ ($\overline{W}$ is given in (\ref{eqns:w}))
gives\begin{equation}
\label{eqns:dtwki}
\left\{
       \begin{array}{ll}
        a_x=\lambda pc-\lambda qb, \\
        b_x=2\lambda b+2hb-2\lambda pa,\\
       c_x=-2\lambda c-2hc+2\lambda qa,\\
              d_x=2\lambda d-2\lambda pf- 2\lambda ra+2hd,\\
              f_x=\lambda p g+\lambda rc- \lambda sb-\lambda qd,\\
      g_x=-2\lambda g+2\lambda qf+2\lambda sa-2gh.\\
       \end{array}
     \right.
\end{equation}
 Further, let $a,b,c$ possess the Laurent expansions:
$$a=\sum_{i\geq 0}a_i\lambda^{-i},~b=\sum_{i\geq 0}b_i\lambda^{-i},~c=\sum_{i\geq 0}c_i\lambda^{-i},
$$$$ f=\sum_{i\geq 0}f_i\lambda^{-i},~d=\sum_{i\geq 0}d_i\lambda^{-i},~g=\sum_{i\geq 0}g_i\lambda^{-i}.
$$
Equivalently,  the system (\ref{eqns:dtwki}) leads to the recursion relations
\begin{equation}
\label{eqns:dt1wki}
\left\{
       \begin{array}{ll}
        a_{ix}= pc_{i+1}- qb_{i+1}, \\
        b_{ix}=2b_{i+1}-2pa_{i+1}+2hb_i,\\
       c_{ix}=-2c_{i+1}+2qa_{i+1}-2hc_i,\\
       d_{ix}=2d_{i+1}-2ra_{i+1}-2pf_{i+1}+2hd_i,~~~~~i\geq 0\\
       f_{ix}=-s b_{i+1}+ rc_{i+1}- q d_{i+1}+pg_{i+1},\\
      g_{ix}=-2g_{i+1}+sa_{i+1}+qf_{i+1}-hg_i.\\
       \end{array}
     \right.
\end{equation}
From (\ref{eqns:dt1wki}), we can find that the initial data satisfy
\begin{equation}
\label{eqns:cz}pc_0-qb_0=0,~b_0=pa_0,~c_0=qa_0,~pg_0+rc_0-sb_0-qd_0=0,~d_0=pf_0+ra_0, ~g_0=qf_0+sa_0,\end{equation}
and $a_{m+1},~e_{m+1}$ have the recursion relations
\begin{equation}
\label{eqns:cz}\begin{array}{ll}
a_{m+1}=\frac{1}{\sqrt{1+pq}}\partial^{-1}\frac{1}{\sqrt{1+pq}}[\frac{1}{4}pc_{mxx}-\frac{1}{4}qb_{mxx}+\frac{1}{2}p(hc_{m})_{x}+\frac{1}{2}q(hb_{m})_{x}-ha_{mx}],\\
f_{m+1}=\frac{1}{\sqrt{1+pq}}\partial^{-1}\frac{1}{\sqrt{1+pq}}[-\sqrt{ps+qr}(\sqrt{ps+qr}a_{m+1})_x+\frac{1}{4}pg_{mxx}+\frac{1}{2}p(hg_{m})_{x}+\frac{1}{4}rc_{mxx}\\
~~~~~~~~~~+\frac{1}{2}r(hc_{m})_{x}-\frac{1}{4}sb_{mxx}+\frac{1}{2}s(hb_{m})_{x}-\frac{1}{4}qf_{mxx}+\frac{1}{2}q(hf_{m})_{x}].
 \end{array}\end{equation}

 To guarantee the uniqueness of
 $\{a_i,b_i,c_i,d_i,f_i,g_i|i\geq 1\}$ , we impose the conditions on constants of integration
 $$a_i\mid_{u=0}=b_i\mid_{u=0}=c_i\mid_{u=0}=d_i\mid_{u=0}=f_i\mid_{u=0}=g_i\mid_{u=0}.$$
 From (\ref{eqns:dt1wki}) to (\ref{eqns:cz}), we  can have
$$a_0=\frac{1}{\sqrt{1+pq}},~b_0=\frac{p}{\sqrt{1+pq}},~c_0=\frac{q}{\sqrt{1+pq}},~f_0=-\frac{ps+qr}{2(1+pq)^{\frac{3}{2}}},~
d_0=\frac{r}{\sqrt{1+pq}}-\frac{p(ps+qr)}{2(1+pq)^{\frac{3}{2}}},$$$$g_0=\frac{s}{\sqrt{1+pq}}-\frac{q(ps+qr)}{2(1+pq)^{\frac{3}{2}}},~
a_1=\frac{1}{4}\frac{4\varepsilon pq(ps+qr)_x+pq_x-qp_x}{(1+pq)^{\frac{2}{3}}},~b_1=\frac{1}{2}\frac{2\varepsilon p(ps+qr)_x+p_x}{(1+pq)^{\frac{2}{3}}},$$$$c_1=-\frac{1}{2}\frac{2\varepsilon q(ps+qr)_x+q_x}{(1+pq)^{\frac{2}{3}}}$$
$$f_1=\frac{1}{8(1+pq)^2}[2pq(r_x-s_x)+ps(3pq_x-qp_x)+qr(pq_x-3qp_x)-2q_xr+2qr_x+2p_xs-ps_x$$
$$+4\varepsilon(ps+qr)_x(ps+qr)(pq-2)],~~~~~~~~~~~~~~~~~~~~~~~~~~~~~~~~~~~~~~$$
$$d_1=(\frac{r}{2\sqrt{1+pq}})_x-(\frac{p(ps+qr)}{2(1+pq)^{\frac{3}{2}}})_x+\frac{r(pq_x-qp_x)}{4(1+pq)^{\frac{2}{3}}}
+\frac{p}{8(1+pq)^2}[2pq(r_x-s_x)+ps(3pq_x-qp_x)+qr(pq_x-3qp_x)$$
$$~~~~~~~~-2q_xr+2qr_x+2p_xs-ps_x+4\varepsilon(ps+qr)_x(ps+qr)(pq-2)]-\varepsilon(\frac{r}{\sqrt{1+pq}}-\frac{p(ps+3qr)}{2(1+pq)^{\frac{3}{2}}})(ps+qr)_x,$$
$$g_1=-(\frac{s}{2\sqrt{1+pq}})_x+(\frac{q(ps+qr)}{4(1+pq)^{\frac{3}{2}}})_x+\frac{s(pq_x-qp_x)}{4(1+pq)^{\frac{2}{3}}}+\frac{q}{8(1+pq)^2}[2pq(r_x-s_x)
+ps(3pq_x-qp_x)+qr(pq_x-3qp_x)$$
$$~~~~~~~~~-2q_xr+2qr_x+2p_xs-ps_x+4\varepsilon(ps+qr)_x(ps+qr)(pq-2)]-\varepsilon(\frac{s}{\sqrt{1+pq}}-\frac{q(3ps+qr)}{2(1+pq)^{\frac{3}{2}}})(ps+qr)_x.$$

Taking
\begin{equation}
\label{eqns:v}
V^{[m]}=\lambda^2(\lambda^{m}W)_++\left[
      \begin{array}{cccc}
       A_m &\lambda B_m &0& \lambda F_m\\
    \lambda C_m& - A_m & \lambda G_m &0 \\
      0&0& A_m& \lambda B_m\\
      0&0&  \lambda C_m & - A_m
      \end{array}
    \right],
\end{equation}
where  $ B_m=\frac{1}{2}(b_{mx}-2hb_m),$
$C_m=\frac{1}{2}(-c_{mx}-2hc_m),~F_m=\frac{1}{2}(d_{mx}-2hd_m)$ and $G_m=\frac{1}{2}(-g_{mx}-2hg_m).$ Thus, the corresponding zero curvature equation
 (\ref{eqns:zce})
\label{eqns:zce}
give rise to a hierarchy
\begin{equation}
\label{eqns:hwki}
\left\{
       \begin{array}{ll}
       p_{t_m}=B_{mx}-2hB_m+2pA_m, \\
       q_{t_m}=C_{mx}-2hC_m-2qA_m,~m\geq 0.\\
        r_{t_m}=F_{mx}-2hF_m+2rA_m,\\
         s_{t_m}=G_{mx}+2hG_m-2sA_m,\\
     h_{t_m}=A_{mx}.
       \end{array}
     \right.
\end{equation}

Based on (\ref{eqns:dt1wki}) and (\ref{eqns:hwki}), we can have
$$A_{mx}=h_{t_m}=\varepsilon (p_{tm}s+ps_{tm}+q_{tm}r+qr_{tm})_x$$
$$=\varepsilon[\frac{1}{2}b_{mxx}s-(hb_m)_xs-\frac{1}{2}pg_{mxx}-(hg_m)_xp-\frac{1}{2}rc_{mxx}-r(hc_m)_x+\frac{1}{2}qd_{mxx}-q(hd_m)_x+2hf_{mx}]_{x}$$
$$=-2\varepsilon ((1+pq)f_{m+1,x}+\frac{1}{2}(pq_x+qp_x)f_{m+1}+(ps+qr)a_{m+1,x}+\frac{1}{2}(ps_x+sp_x+rq_x+qr_x)a_m+1)_x$$
$$=-2\varepsilon[\sqrt{1+pq}(\sqrt{1+pq}f_{m+1})_x+\sqrt{ps+qr}(\sqrt{ps+qr}a_{m+1})_x]_x$$
So, we can choose $A_m=-2\varepsilon[\sqrt{1+pq}(\sqrt{1+pq}f_{m+1})_{x}+\sqrt{ps+qr}(\sqrt{ps+qr}a_{m+1})_x],$ and generate the complete system of the WKI integrable couplings:
\begin{equation}
\label{eqns:cs}\left[
      \begin{array}{c}
       p_{tm}\\
    q_{tm} \\
     r_{tm}\\
    s_{tm}
      \end{array}
    \right]=\left[
      \begin{array}{c}
       \frac{1}{2}b_{mxx}-(hb_m)_x-h(b_{mx}-2hb_m)-4p\varepsilon[\sqrt{1+pq}(\sqrt{1+pq}f_{m+1})_x+\sqrt{ps+qr}(\sqrt{ps+qr}a_{m+1})_x)]\\
     -\frac{1}{2}c_{mxx}-(hc_m)_x-h(c_{mx}+2hc_m)+4q\varepsilon[\sqrt{1+pq}(\sqrt{1+pq}f_{m+1})_x+\sqrt{ps+qr}(\sqrt{ps+qr}a_{m+1})_x)] \\
      \frac{1}{2}d_{mxx}-(hd_m)_x-h(d_{mx}-2hd_m)-4r\varepsilon[\sqrt{1+pq}(\sqrt{1+pq}f_{m+1})_x+\sqrt{ps+qr}(\sqrt{ps+qr}a_{m+1})_x)]\\
      -\frac{1}{2}g_{mxx}-(hg_m)_x-h(g_{mx}-2hg_m)+4s\varepsilon[\sqrt{1+pq}(\sqrt{1+pq}f_{m+1})_x+\sqrt{ps+qr}(\sqrt{ps+qr}a_{m+1})_x)]
      \end{array}
    \right],~~~~~m\geq 0.
\end{equation}
The first nonlinear equation in the above new system can be given
$$p_{t_0}=(\frac{p}{2\sqrt{1+pq}})_{xx}-\varepsilon \frac{p(ps+qr)_{xx}}{\sqrt{1+pq}}-2\varepsilon (ps+qr)_{x}(\frac{p}{\sqrt{1+pq}})_x+\frac{2\varepsilon^2 p(ps+qr)_{x}^2}{\sqrt{1+pq}}-\varepsilon p\sqrt{1+pq}[\frac{1}{2(1+pq)^{\frac{3}{2}}}$$
$$~~~~~~~~~~~~(2pq(r_x-s_x)+ps(3pq_x-qp_x)+qr(pq_x-3qp_x)-2q_xr+2qr_x+2p_xs-ps_x+4\varepsilon(ps+qr)_x(ps+qr)(pq-2))]_x$$
$$-4\varepsilon p\sqrt{ps+qr}[\frac{\sqrt{ps+qr}(4\varepsilon pq(ps+qr)_x+pq_x-qp_x)}{4(1+pq)^{\frac{2}{3}}}]_x,~~~~~~~~~~~~~~~~~~~~~~~~~~~~~~~~~~~~$$
$$q_{t_0}=-(\frac{q}{\sqrt{1+pq}})_{xx}-\varepsilon \frac{q(ps+qr)_{xx}}{\sqrt{1+pq}}-2\varepsilon (ps+qr)_{x}(\frac{q}{\sqrt{1+pq}})_x-\frac{2\varepsilon^2 q(ps+qr)_{x}^2}{\sqrt{1+pq}}-\varepsilon q\sqrt{1+pq}[\frac{1}{2(1+pq)^{\frac{3}{2}}}$$
$$~~~~~~~~~~~~(2pq(r_x-s_x)+ps(3pq_x-qp_x)+qr(pq_x-3qp_x)-2q_xr+2qr_x+2p_xs-ps_x+4\varepsilon(ps+qr)_x(ps+qr)(pq-2))]_x$$
$$-4\varepsilon q\sqrt{ps+qr}[\frac{\sqrt{ps+qr}(4\varepsilon pq(ps+qr)_x+pq_x-qp_x)}{4(1+pq)^{\frac{2}{3}}}]_x,~~~~~~~~~~~~~~~~~~~~~~~~~~~~~~~~~~~~$$
$$r_{t_0}=(\frac{r}{2\sqrt{1+pq}}-\frac{p(ps+qr)}{4(1+pq)^{\frac{3}{2}}})_{xx}-\varepsilon (ps+qr)_{xx}[\frac{r}{\sqrt{1+pq}}-\frac{p(ps+qr)}{2(1+pq)^{\frac{3}{2}}}]-2\varepsilon (ps+qr)_{x}[\frac{r}{\sqrt{1+pq}}-\frac{p(ps+qr)}{2(1+pq)^{\frac{3}{2}}}]_x$$
$$+2\varepsilon^2(ps+qr)_{x}^{2}[\frac{r}{\sqrt{1+pq}}-\frac{p(ps+qr)}{2(1+pq)^{\frac{3}{2}}}]-\varepsilon r\sqrt{1+pq}[\frac{1}{2(1+pq)^{\frac{3}{2}}}(2pq(r_x-s_x)+ps(3pq_x-qp_x)+qr(pq_x-3qp_x)$$
$$-2q_xr+2qr_x+2p_xs-ps_x+4\varepsilon(ps+qr)_x(ps+qr)(pq-2))]_x-4\varepsilon r\sqrt{ps+qr}[\frac{\sqrt{ps+qr}(4\varepsilon pq(ps+qr)_x+pq_x-qp_x)}{4(1+pq)^{\frac{2}{3}}}]_x,$$
$$s_{t_0}=-(\frac{s}{2\sqrt{1+pq}}-\frac{q(ps+qr)}{4(1+pq)^{\frac{3}{2}}})_{xx}-\varepsilon (ps+qr)_{xx}[\frac{s}{\sqrt{1+pq}}-\frac{q(ps+qr)}{2(1+pq)^{\frac{3}{2}}}]-2\varepsilon (ps+qr)_{x}[\frac{s}{\sqrt{1+pq}}-\frac{q(ps+qr)}{2(1+pq)^{\frac{3}{2}}}]_x$$
$$+2\varepsilon^2(ps+qr)_{x}^{2}[\frac{s}{\sqrt{1+pq}}-\frac{q(ps+qr)}{2(1+pq)^{\frac{3}{2}}}]-\varepsilon s\sqrt{1+pq}[\frac{1}{2(1+pq)^{\frac{3}{2}}}(2pq(r_x-s_x)+ps(3pq_x-qp_x)+qr(pq_x-3qp_x)$$
$$-2q_xr+2qr_x+2p_xs-ps_x+4\varepsilon(ps+qr)_x(ps+qr)(pq-2))]_x-4\varepsilon s\sqrt{ps+qr}[\frac{\sqrt{ps+qr}(4\varepsilon pq(ps+qr)_x+pq_x-qp_x)}{4(1+pq)^{\frac{2}{3}}}]_x.$$

\subsection{Completion of the vector  AKNS integrable couplings}
We consider the matrix iso-spectral problem
\begin{equation}
\label{eqns:0a}
\phi_x=\overline{U}\phi,~\bar{U}=\left[
      \begin{array}{cc}
        U &U_1\\
      0 &U \\
      \end{array}
    \right]=\left[
      \begin{array}{cc|cc}
       \lambda +h & p &0&  r\\
       q & -(\lambda+h) I_N& s &0 \\ \hline
      0&0& \lambda +h &p\\
      0&0&  q & -(\lambda+h) I_N
      \end{array}
    \right],~~~h=\varepsilon(ps+rq),
\end{equation}
where $p=(p_1,p_2,\cdots,p_N),~q=(q_1,q_2,\cdots, q_N)^{T},~ r=(r_1,r_2,\cdots,r_N),s=(s_1,s_2,\cdots,s_N)^{T}$, $I_N$ is a $n-$ order unit matrix.
 Here, a nonlinear perturbation term $h$ is added to the spectral matrix. When $\varepsilon=0$, it reduces the case of vector AKNS integrable couplings\cite{akns1,akns2}. Based on  thus spectral matrix, we can work out the  complete system of the vector AKNS integrable couplings by standard procedure.

Assume
\begin{equation}
\label{eqns:wAKNS}
\overline{W}=\left[
      \begin{array}{cc}
        W &W_1\\
      0 &W \\
      \end{array}
    \right]=\left[
      \begin{array}{cc|cc}
       a &b &d&e\\
      c & -aI_N &f&-dI_N \\ \hline
      0&0& a&b\\
      0&0&c & -aI_N
      \end{array}
    \right]
\end{equation}where $b,~e$ are row vectors, $c, ~f$ are column vectors, and $a,~d$ are real numbers.

Solving the stationary zero curvature equation
$
\overline{W}_x=[\overline{U},\overline{W}]
$
gives\begin{equation}
\label{eqns:dta}
\left\{
       \begin{array}{ll}
        a_x=pc-bq, \\
        b_x=2(\lambda+h) b-2a p,\\
       c_x=-2(\lambda+h) c+2a q,\\
              d_x=pf+rc-bs-eq,\\
              e_x=2(\lambda+h) e-2dp-2ar,\\
      f_x=-2(\lambda+h)f+2dq+2as.\\
       \end{array}
     \right.
\end{equation}
Let us seek a formal solution of the type
\begin{equation}
\label{eqns:wAKNSzhank}
\overline{W}=\left[
      \begin{array}{cc|cc}
       a &b &d&e\\
      c & -aI_N &f&-dI_N \\ \hline
      0&0& a&b\\
      0&0&c & -aI_N
      \end{array}
    \right]=\sum_{i\geq 0}\overline{W}_i\lambda^{-i},~\overline{W}_i=\left[
      \begin{array}{cc|cc}
       a_i &b_i &d_i&e_i\\
      c_i & -a_iI_N &f_i&-d_iI_N \\ \hline
      0&0& a_i&b_i\\
      0&0&c_i & -a_iI_N
      \end{array}
    \right]
\end{equation}
 The Eq. (\ref{eqns:dta}) recursively define all $\{a_i,b_i,c_i,d_i,f_i,g_i|i\geq 1\}$ with the initial values
 $$a_0=1,~b_0=\underbrace{(0,0,\cdots,0)}_{N},~c_0=\underbrace{(0,0,\cdots,0)^{T}}_{N},~d_0=1,~e_0=\underbrace{(0,0,\cdots,0)}_{N},~f_0=\underbrace{(0,0,\cdots,0)^{T}}_{N}.$$

For any integer $m\geq1$, we introduce
\begin{equation}
\label{eqns:va}
V^{[m]}=(\lambda^{m}W)_++\triangle_m=\sum_{i=0}^{m}V_i+\triangle_m,~
\triangle_m=\left[
      \begin{array}{cccc}
       \delta_m &0 &0& 0\\
    0& - \delta_m I_N& 0&0 \\
      0&0& \delta_m& 0\\
      0&0&  0 & - \delta_m I_N
      \end{array}
    \right],
\end{equation}
where  $ \delta_m$ is undetermined  function. Thus, the corresponding zero curvature equation
 (\ref{eqns:zce}) with $\overline{U},~V$ defined in (\ref{eqns:0a}) and (\ref{eqns:va})
give rise to a hierarchy
\begin{equation}
\label{eqns:hakns}
\left\{
       \begin{array}{ll}
       p_{t_m}=2b_{m+1}+2\delta_mp, \\
       q_{t_m}=-2c_{m+1}-2\delta_mq,~m\geq 0.\\
        r_{t_m}=2e_{m+1}+2\delta_mr,\\
         s_{t_m}=-2f_{m+1}-2\delta_ms,\\
     h_{t_m}=\delta_{mx}.
       \end{array}
     \right.
\end{equation}

Based on(\ref{eqns:dta}) and  (\ref{eqns:hakns}), we can have
$$\delta_{mx}=h_{t_m}=\varepsilon (p_{tm}s+ps_{tm}+q_{tm}r+qr_{tm})$$
$$=\varepsilon(2b_{m+1}s-2pf_{m+1}+2e_{m+1}q-2rc_{m+1})=-2\varepsilon d_{m+1,x}.$$
So, we can choose $\delta_m=-2\varepsilon d_{m+1}.$  The complete system of the vector AKNS integrable couplings are generated
\begin{equation}
\label{eqns:cs}\left[
      \begin{array}{c}
       p^{T}_{tm}\\
    q_{tm} \\
     r^{T}_{tm}\\
    s_{tm}
      \end{array}
    \right]=\left[
      \begin{array}{c}
       2b^{T}_{m+1}-4\varepsilon p^{T}d_{m+1}\\
     -2c_{m+1}+4\varepsilon qd_{m+1} \\
      2e^{T}_{m+1}-4\varepsilon r^{T}d_{m+1}\\
    -2f_{m+1}+4\varepsilon sd_{m+1}
      \end{array}
    \right],~~~~~m\geq 0.
\end{equation}
The first nonlinear equation in the above new system is
$$p_{t_2}=\frac{1}{2}p_{xx}-(pq)p-\varepsilon[(pq)r_x+(ps)s_x+3(ps)p_x+2(p_xq)r+(pq_x)r+(pq_x-p_xq-p_xs-r_xq+rq_x)p]$$
$$+2\varepsilon^2[(ps+rq)^2p-2(ps+rq)(pq+ps+rq)p],~~~~~~~~~~~~~~~~~~~~~~~~~~~~~~~~~~~~~~~~~~~~~~~~$$
$$q_{t_2}=-\frac{1}{2}q_{xx}+(pq)q-\varepsilon[(r_xq)q+(pq)s_x+(p_xq)s+2(pq_x)s+(rq)q_x+(pq_x-p_xq-p_xs-r_xq+rq_x)q]$$
$$-2\varepsilon^2[(ps+rq)^2q+2(ps+rq)(pq+ps+rq)q],~~~~~~~~~~~~~~~~~~~~~~~~~~~~~~~~~~~~~~~~~~~~~~~$$
$$r_{t_2}=\frac{1}{2}(p_{xx}+r_{xx})-[pq+ps]p-(pq)r-\varepsilon[(rs+rs_x+3ps)p_x+(3rq+2ps+pq)r_x+(pq_x)r
+(ps_x)p$$$$~~~~~+(rq_x+2p_xq
+pq_x-p_xq-p_xs-r_xq+rq_x)r]+2\varepsilon^2[(ps+rq)^2(p+r)-2(ps+rq)(pq+ps+rq)p],$$
$$s_{t_2}=-\frac{1}{2}(q_{xx}+s_{xx})-(pq+rq+ps)q-\varepsilon[(ps_x+r_xq+r_xs+p_xs)q+(rs+ps+3rq)q_x+(3ps+rq)s_x$$$$
~~~~~+(-pq_x+p_xq+p_xs+r_xq+rq_x+p_xs)s]-2\varepsilon^2[(ps+rq)^2(q+s)-2(ps+rq)(pq+ps+rq)s].$$

\section{Completion of the discrete integrable couplings}
In this section, we will show that the procedure of constructing the completion of the integrable couplings  can be applied to the  discrete systems.
In the following, we will take the Volterra lattice hierarchy as an example to illustrate it.

\subsection{Completion of the Volterra integrable couplings}
A generalized discrete spectral problem for the Volterra lattice hierarchy  is given by
\begin{equation}
\label{eqns:0v}
E\phi=\overline{U}\phi,~\bar{U}=\left[
      \begin{array}{cc}
        U &U_1\\
      0 &U \\
      \end{array}
    \right]=\left[
      \begin{array}{cc|cc}
       1& u+h &0&  r\\
       \lambda^{-1}& 0& 0 &0 \\ \hline
      0&0& 1 &\lambda +h\\
      0&0&   \lambda^{-1} & 0
      \end{array}
    \right],~~~h=\varepsilon ur,
\end{equation}
where $E$ is the shift operator and  $(E^{m}f)(n)=f^{m}(n)=f(n+m)$.
 Here, a nonlinear perturbation term $h$ is added to the spectral matrix. When $\varepsilon=0$, it reduces the  integrable couplings of Volterra lattice hierarchy \cite{c6,v1}. Based on  this  spectral matrix, we can work out the  complete system of the Volterra lattice  integrable couplings.

Upon setting
\begin{equation}
\label{eqns:wv}
\overline{\Gamma}=\left[
      \begin{array}{cc}
       \Gamma &\Gamma_1\\
      0 &\Gamma \\
      \end{array}
    \right]=\left[
      \begin{array}{cc|cc}
       a &b &d&f\\
      c & -a &g&-d \\ \hline
      0&0& a&b\\
      0&0&c & -a
      \end{array}
    \right]
\end{equation}
The discrete stationary zero curvature equation
$
(E\overline{\Gamma})\overline{U}-\overline{U}\overline{\Gamma}=0
$
gives\begin{equation}
\label{eqns:dtv}
\left\{
       \begin{array}{ll}
       a^{(1)}-a+\lambda^{-1}b^{(1)}-c(h+u)=0, \\
       (h+u)(a^{(1)}+a)-b=0,\\
       c^{(1)}-\lambda^{-1}(a^{(1)}+a)=0,\\
              (h+u)c^{(1)}-\lambda^{-1} b=0,\\
             -(h+u)g-cr+d^{(1)}-d+\lambda^{-1}f^{(1)}=0,\\
    g^{(1)}-\lambda^{-1}(d^{(1)}+d)=0,\\
   r (a^{(1)}+a)+(u+h)(d^{(1)}+d)-f=0,\\
   r c^{(1)}+(u+h)g^{(1)}-\lambda^{-1}f=0.
       \end{array}
     \right.
\end{equation}
Let $\overline{\Gamma}$ possess the Laurent expansions
\begin{equation}
\label{eqns:wvzhan}
\overline{\Gamma}=\sum_{i\geq 0}\overline{\Gamma}_i\lambda^{-i},~\overline{\Gamma}_i=\left[
      \begin{array}{cc|cc}
       a_i &b_i &d_i&f_i\\
      c_i & -a_i &g_i&-d_i\\ \hline
      0&0& a_i&b_i\\
      0&0&c_i & -a_i
      \end{array}
    \right]
\end{equation}
 The Eq. (\ref{eqns:dtv}) equivalently leads to
 \begin{equation}
\label{eqns:dtvzhan}
\left\{
       \begin{array}{ll}
       a^{(1)}_{i+1}-a_{i+1}-(h+u)c_{i+1}+b^{(1)}_i=0, \\
       b_i=(h+u)(a^{(1)}_i+a_i),\\
       c^{(1)}_{i+1}=a^{(1)}_i+a_i,\\
              (h+u)c^{(1)}_{i+1}= b_i,\\
             -(h+u)g_{i+1}-rc_{i+1}+d^{(1)}_{i+1}-d_{i+1}=-f^{(1)}_i,\\
     g^{(1)}_{i+1}=d^{(1)}_i+d_i,\\
     r(a^{(1)}_i+a_i)+(u+h)(d^{(1)}_i+d_i)-f_i=0,\\
     rc^{(1)}_{i+1}+(h+u)g^{(1)}_{i+1}-f_i=0
       \end{array}
     \right.
\end{equation}
 Upon choosing $a_0=\frac{1}{2},~e_0=\frac{1}{2},~a_i\mid_{u=0}=0,~e_i\mid_{u=0}=0,~i\geq 1$,  all sets of the functions $a_i,~b_i,~c_i,~d_i,~f_i$ and $g_i$ can be uniquely determined. The first two sets are
  $$a_0=\frac{1}{2},~b_0=u+h,~c_0=0,~d_0=\frac{1}{2},~f_0=h+u+r,~g_0=0,$$
$$a_1=-(h+u),~c_1=1,~b_1=-(h+u)(u^{(1)}+u),~d_1=-(h+u+r),~g_1=1,$$
$$f_1=-r(u^{(1)}+u)-(h+u)(h^{(1)}+u^{(1)}+r^{(1)}+h+u+r)$$
The compatibility conditions  of the matrix discrete spectral problem
$$E\phi=\overline{U}\phi,~\phi_t=V^{[m]}\phi,~V^{[m]}=(\lambda^{m+1}\Gamma)_{+}+\Delta_m,~\Delta_m=\left[
      \begin{array}{cccc}
       0 &-b_{m+1} &0& -f_{m+1}\\
    0& a_{m+1}+a_{m+1}^{(-1)}& 0&d_{m+1}+d_{m+1}^{(-1)} \\
      0&0& 0& -b_{m+1}\\
      0&0&  0 & a_{m+1}+a_{m+1}^{(-1)}
      \end{array}
    \right]$$
determine the
the  complete system of the Volterra lattice  integrable couplings
\begin{equation}
\label{eqns:cs}\left[
      \begin{array}{c}
       u_{tm}\\
    r_{tm}
           \end{array}
    \right]=\left[
      \begin{array}{c}
      \frac{1}{1+\varepsilon r} (-u(a_{m+1}+a_{m+1}^{(-1)})+\varepsilon u^2(\varepsilon r+1)(d_{m+1}+d_{m+1}^{(-1)})+b_{m+1}-\varepsilon uf_{m+1})\\
        -(\varepsilon r+1)u(d_{m+1}+d_{m+1}^{(-1)})-r(a_{m+1}+a_{m+1}^{(-1)})+f_{m+1}
               \end{array}
    \right]
\end{equation}
The first nonlinear equation in the above new system is
$$u_{t_0}=\frac{1}{1+\varepsilon r}[(1+\varepsilon r^{(-1)})uu^{(-1)}-uu^{(1)}+\varepsilon u^2 r],~~~~~~~~~~~~~~~~~~~~~~~~~~~~~~~~~~~~~~~~~~~~~~~~~~~~~~~~~~~~~~$$
$$r_{t_0}=u(E+E^{-1})(u+r)+\varepsilon [uu^{(-1)}(1+E^{-1})r+ur(E^{-1}-E+1)r+u^{(1)}(rr^{(1)}-ur-ur^{(1)})]+\varepsilon^{2}ur(E^{-1}+E)ur.$$

\section{Completion of the super integrable couplings}
In this section, we will take super Dirac and super NLS-mKdV integrable hierarchies  to show that the procedure of constructing the completion of the integrable couplings  can be applied to the  super integrable system\cite{s1,s2,s3}.

\subsection{Completion of super Dirac integrable couplings}
Now we consider the following super  spectral problem
\begin{equation}
\label{eqns:0d}
\phi_x=\overline{U}\phi,~\bar{U}=\left[
      \begin{array}{ccc|ccc}
       p& \lambda +h +q &\alpha &  r&s&\xi\\
       -\lambda -h +q & -p& \beta &s&-r&\eta\\
      \beta&-\alpha& 0 &\eta &-\xi&0\\ \hline
      0&0&  0&  p& \lambda +h +q &\alpha\\
       0&0&  0& -\lambda -h +q & -p& \beta \\
         0&0&  0&\beta &-\alpha & 0
      \end{array}
    \right],~~~h=\varepsilon(qs+pr+\alpha\eta+\xi\beta).
\end{equation}
Here we would like to point out that $p,~q,~r,~s,~\alpha,~\beta,~\xi$  and $\eta$ are all functions of $x$ and $t$. $p,~q,~r,~s$ and $\lambda$ are
bosonic, and $\alpha,~\beta,~\xi$  and $\eta$ are fermionic.
In the spectral problem (\ref{eqns:0d}), a nonlinear perturbation term $h$ is added. When $\varepsilon=0$, it reduces the case of super Dirac integrable couplings \cite{sd1}.

Upon setting
\begin{equation}
\label{eqns:wd}
\overline{W}=\left[
      \begin{array}{ccc|ccc}
       c &a+b &\rho &d&f+g& \mu\\
      a-b & -c &\sigma &f-g & -d &\nu \\
     \sigma&-\rho &0 &\nu &-\mu & 0\\ \hline
      0&0&0 & c &a+b &\rho\\
      0&0&0 &a-b & -c &\sigma \\
       0&0&0&\sigma&-\rho &0
      \end{array}
    \right].
\end{equation}

Solving the stationary zero curvature equation
$
\overline{W}_x=[\overline{U},\overline{W}]
$
gives\begin{equation}
\label{eqns:dtd}
\left\{
       \begin{array}{ll}
        a_x=-2\lambda c-2hc+2pb-\alpha \rho+\beta \sigma, \\
        b_x=2pa-2qc-\alpha \rho-\beta \sigma,\\
       c_x=2\lambda a-2qb+2ha+\alpha \sigma+\beta \rho,\\
             \rho_x=(\lambda+q)\sigma-\beta(a+b)-\alpha c+p \rho+h\sigma,\\
              \sigma_x=(-\lambda+q)\rho-p\sigma-\alpha (a-b)+\beta c-h\rho,\\
      d_x=2\lambda f+2hf-2q g-2bs +\alpha \gamma+\xi \sigma-\rho \eta-\mu \beta,\\
      f_x=-2\lambda d-2hd+2pg+2rb-\alpha \mu-\xi \rho+\eta \sigma+\beta \nu,\\
      g_x=2ra-2qd+2pf-2sc-\alpha \mu-\xi \rho-\eta \sigma-\beta \nu,\\
      \mu_x=\lambda \nu+p\mu+(h+q)\nu+r\rho+s\sigma-\beta (f+g)-\alpha d-\eta (a+b)-\xi c,\\
      \nu_x=-\lambda \mu+(q-h)\mu-p\nu+s\rho-r\sigma+\beta d+\alpha (f+g)-\xi (a-b)+\eta c.
       \end{array}
     \right.
\end{equation}
Let us seek a formal solution of the type
\begin{equation}
\label{eqns:wAKNSzhank}
\overline{W}=\sum_{i\geq 0}\overline{W}_i\lambda^{-i},~\overline{W}_i=\left[
      \begin{array}{ccc|ccc}
       c_i &a_i+b_i &\rho_i &d_i&f_i+g_i& \mu_i\\
      a_i-b_i & -c_i &\sigma_i &f_i-g_i & -d_i &\nu_i \\
     \sigma_i&-\rho_i &0 &\nu_i &-\mu_i & 0\\ \hline
      0&0&0 & c_i &a_i+b_i &\rho_i\\
      0&0&0 &a_i-b_i & -c_i &\sigma_i \\
       0&0&0&\sigma_i&-\rho_i &0
      \end{array}
    \right]
\end{equation}
 The Eq. (\ref{eqns:dtd}) equivalently leads to
\begin{equation}
\label{eqns:dtdz}
\left\{
       \begin{array}{ll}

       a_{i+1}=\frac{1}{2} c_{ix}+qb_i-ha_i-\frac{1}{2}\alpha \sigma_i-\frac{1}{2}\beta \rho_i,\\
              b_{i+1,x}=2pa_{i+1}-2qc_{i+1}-\alpha \rho_{i+1}-\beta \sigma_{i+1},\\
       c_{i+1}=-\frac{1}{2}a_{ix}-hc_i+pb_i-\frac{1}{2}\alpha \rho_i+\frac{1}{2}\beta \sigma_i,\\
             \sigma_{i+1}=\frac{1}{2}\rho_{ix}-q\sigma_i+\beta(a_i+b_i)+\alpha c_i-p \rho_i-h\sigma_i,\\
             \rho_{i+1}=-\frac{1}{2}\sigma_x +q\rho_i-p\sigma_i-\alpha (a_i-b_i)+\beta c_i-h\rho_i,\\
    f_{i+1}= \frac{1}{2}  d_x-hf_i+q g_i+sb_i -\frac{1}{2} \alpha \gamma_i-\frac{1}{2} \xi \sigma_i+\frac{1}{2} \rho_i \eta+\frac{1}{2} \mu_i \beta,\\
     d_{i+1}=-\frac{1}{2} f_{ix}-hd_i+pg_i+rb_i-\frac{1}{2}\alpha \mu_i-\frac{1}{2}\xi \rho_i+\frac{1}{2}\eta \sigma_i+\frac{1}{2}\beta \nu_i,\\
      g_{i+1,x}=2ra_{i+1}-2qd_{i+1}+2pf_{i+1}-2sc_{i+1}-\alpha \mu_{i+1}-\xi \rho_{i+1}-\eta \sigma_{i+1}-\beta \nu_{i+1},\\
       \nu_{i+1}=\mu_{ix}-p\mu_i-(h+q)\nu_i-r\rho_i-s\sigma_i+\beta (f_i+g_i)+\alpha d_i+\eta (a_i+b_i)+\xi c_i,\\
     \mu_{i+1}=- \nu_{ix}+(q-h)\mu_i-p\nu_i+s\rho_i-r\sigma_i+\beta d_i-\alpha (f_i+g_i)-\xi (a_i-b_i)+\eta c_i.
       \end{array}
     \right.
\end{equation}

 Choosing the initial values as
 $$a_0=c_0=\rho_0=\sigma_0=0,~b_i=1,~d_0=f_0=\mu_0=\nu_0=0,~g_0=1$$
 and assuming $a_i\mid_{u=0}=b_i\mid_{u=0}=c_i\mid_{u=0}=d_i\mid_{u=0}=f_i\mid_{u=0}=g_i\mid_{u=0}\rho_i\mid_{u=0}=\sigma_i\mid_{u=0}
 =\mu_i\mid_{u=0}=\nu_i\mid_{u=0}=0,~i\geq 1$, the recursion relation (\ref{eqns:dtdz}) uniquely defines all differential polynomial  functions
 $a_i,~b_i,~c_i,~d_i,~f_i,~g_i,~\rho_i,~\sigma_i,~\mu_i,$ and $\nu_i$. The first two sets are:
  $$a_1=q,~b_1=0~c_1=p,~\rho_1=\alpha,~\sigma_1=\beta,~d_1=p+r,~f_1=q+s,~g_1=0,~\mu_1=\xi+\alpha,~\nu_1=\eta+\beta,$$
  $$a_2=\frac{1}{2}p_x-\varepsilon q(qs+pr+\alpha \eta+\xi \beta),~b_2=\frac{1}{2}(q^2+p^2+\alpha \beta),~c_2=-\frac{1}{2}q_x-\varepsilon p(qs+pr+\alpha \eta+\xi \beta)$$
 $$\sigma_2=\alpha_x-\varepsilon \beta(qs+pr+\alpha \eta+\xi \beta),~\rho_2=-\beta_x-\varepsilon \alpha(qs+pr+\alpha \eta+\xi \beta),~~~~~~~~~~~~~~~~~~~~~~~~~~~~~~~~~$$
 $$d_2=-\frac{1}{2}(q_x+s_x)-\varepsilon (qs+pr+\alpha \eta+\xi \beta)(p+r),~f_2=\frac{1}{2}(p_x+r_x)-\varepsilon (qs+pr+\alpha \eta+\xi \beta)(q+s),$$
 $$\mu_2=-\eta_x-\beta_x-\varepsilon (qs+pr+\alpha \eta+\xi \beta)(\alpha+\xi),~\nu_2=\xi_x+\alpha_x-\varepsilon (qs+pr+\alpha \eta+\xi \beta)(\eta+\beta),~~~~~~~$$
 $$g_2=pr+qs+\frac{1}{2}(p^2+q^2)+\alpha \eta+(\alpha+\xi) \beta.~~~~~~~~~~~~~~~~~~~~~~~~~~~~~~~~~~~~~~~~~~~~~~~~~~~~~~~~~~~~~~~~~~~~~$$

For any integer $m\geq1$, we take
\begin{equation}
\label{eqns:vad}
V^{[m]}=(\lambda^{m}W)_++\triangle_m=\sum_{i=0}^{m}V_i+\triangle_m,~
\triangle_m=\left[
      \begin{array}{cccccc}
       0 &\delta_m &0&0&0& 0\\
     -\delta_m & 0 &0&0&0&0 \\
     0&0&0& 0&0&0\\
      0&0&0 & 0 &\delta_m  &0\\
      0&0&0 &-\delta_m &0 &0 \\
       0&0&0&0&0 &0
      \end{array}
    \right]
\end{equation}
where  $ \delta_m$ is undetermined  function. Thus, the corresponding zero curvature equation
 (\ref{eqns:zce}) with $\overline{U},~V$ defined in (\ref{eqns:0d}) and (\ref{eqns:vad})
yields
\begin{equation}
\label{eqns:hd}
\left\{
       \begin{array}{ll}
       p_{t_m}=2a_{m+1}+2q\delta_m, \\
       q_{t_m}=-2c_{m+1}-2p\delta_m,~m\geq 0.\\
        r_{t_m}=2f_{m+1}+2s\delta_m,\\
         s_{t_m}=-2d_{m+1}-2r\delta_m,\\
    \alpha_{t_m}=\sigma_{m+1}+\beta\delta_m,\\
   \beta_{t_m}=-\rho_{m+1}-\alpha\delta_m,\\
   \xi_{t_m}=\gamma_{m+1}+\eta\delta_m,\\
   \eta_{t_m}=-\mu_{m+1}-\xi\delta_m,\\
     h_{t_m}=\delta_{mx}.
       \end{array}
     \right.
\end{equation}

Based on (\ref{eqns:dtd}) and (\ref{eqns:hd}), we can have
$$\delta_{mx}=h_{t_m}=\varepsilon (q_{tm}s+qs_{tm}+p_{tm}r+pr_{tm}+\alpha \eta_{tm}+\alpha_{tm}\eta+\xi \beta_{tm}+\xi_{tm}\beta)$$
$$=\varepsilon(-2sc_{m+1}s-2qd_{m+1}+2ra_{m+1}q+2pf_{m+1}+\sigma_{n+1}\eta-\alpha\mu_{n+1}+\gamma_{n+1}\beta-\xi\rho_{n+1})=\varepsilon g_{m+1,x}.$$
So, we can choose $\delta_m=\varepsilon g_{m+1}.$  The complete system of the super dirac  integrable couplings are generated
\begin{equation}
\label{eqns:cs}\left[
      \begin{array}{c}
       p_{tm}\\
    q_{tm} \\
     r_{tm}\\
    s_{tm}\\
    \alpha_{t_m}\\
   \beta_{t_m}\\
   \xi_{t_m}\\
   \eta_{t_m}
      \end{array}
    \right]=\left[
      \begin{array}{c}
       2a_{m+1}+2\varepsilon q g_{m+1} \\
       -2c_{m+1}-2\varepsilon pg_{m+1}\\
       2f_{m+1}+2\varepsilon s g_{m+1}\\
         -2d_{m+1}-2\varepsilon r g_{m+1} \\
   \sigma_{m+1}+\varepsilon   \beta g_{m+1}\\
   -\rho_{m+1}-\varepsilon \alpha g_{m+1}\\
   \gamma_{m+1}+\varepsilon \eta g_{m+1}\\
   -\mu_{m+1}-\varepsilon \xi g_{m+1}\\
      \end{array}
    \right],~~~~~m\geq 0.
\end{equation}
The first nonlinear super system is
$$p_{t_2}=-\frac{1}{2} q_{xx}-\frac{1}{2}(\alpha\alpha_x-\beta\beta_x)+q(q^2+p^2+\alpha \beta)+\varepsilon[2q(\alpha\alpha_x+\beta\beta_x+\eta\eta_x+\xi\xi_x-pq_x+sr_x$$$$~~~~~~~-rs_x-sp_x+qp_x)-2rpp_x-2p_x(\alpha\eta+\xi\beta)]+4\varepsilon^2[q^2s(
pr+\alpha\eta+\xi\beta)+2q(p^2r^2+q^2s^2$$
$$+2\alpha\eta\xi\beta)+2pqr(\alpha\eta+\xi\beta)]+2\varepsilon^2[(ps+rq)^2p-2(ps+rq)(pq+ps+rq)p],~~~~~~~$$
$$q_{t_2}=\frac{1}{2} p_{xx}-\frac{1}{2}(\alpha\beta_x+\beta\alpha_x)-p(q^2+p^2+\alpha \beta)-2\varepsilon[p(\alpha\alpha_x+\beta\beta_x+\eta\eta_x+\xi\xi_x-qp_x+sr_x$$$$-rs_x-pq_x+rq_x)+q_x(sq+\alpha\eta+\xi\beta)]-4\varepsilon^2[p^2r(
pr+qs+\alpha\eta+\xi\beta)~~~~~~~~~~~~~$$$$+pqs(\beta\xi+\alpha\eta+\frac{1}{2}qs)+p\alpha\eta\xi\beta)],~~~~~~~~~~~~~~~~~~~~~~~~~~~~~~~~~~~~~~~~~~~~~~~~~~~~~~~~~~$$
$$r_{t_2}=-\frac{1}{2} (q_{xx}+s_{xx})-\frac{1}{2}(\xi\alpha_x+\eta\beta_x)-\alpha(\xi_x+\alpha_x)-\beta(\beta_x+\eta_x)+2q^2(r+s+q)+p^2(q+s)+2q(\alpha\beta+ps+\alpha\eta+\beta\xi)$$
$$~~~~~~~~~+s\alpha\beta+2\varepsilon[s(-pq_x+\alpha\alpha_x+\beta\beta_x+\xi\xi_x+\eta\eta_x-rs_x-qr_x)-p_x(\alpha\eta+\xi\beta+pr)+r_x(\alpha\eta+\xi\beta+s^2)]$$
$$~~~~+2\varepsilon^2[
q^2s(s^2+qs+\beta\xi+\alpha\eta+pr)+s(\alpha\eta+\beta\xi+spr)(pr+qs)+\alpha\beta\eta\xi(q+s)+\beta\xi pqr],$$
$$s_{t_2}=\frac{1}{2} (p_{xx}+r_{xx})-\frac{1}{2}(\xi\beta_x+\eta\alpha_x)-\beta(\xi_x+\alpha_x)-\alpha(\beta_x+\eta_x)-q^2(r+p)-p^2(p+r+2s)-2p(qr+\alpha\eta+\beta\xi+2\alpha\beta)$$
$$~~~~~~~~~-r\alpha\beta+2\varepsilon[r(\alpha\alpha_x+\beta\beta_x+\xi\xi_x+\eta\eta_x+qp_x+sr_x-qr_x)-q_x(\alpha\eta+\xi\beta+qs)-s_x(\alpha\eta+\xi\beta+pr+qs-r^2)]$$
$$~~~~~~~~~(\eta\beta+\alpha\xi)(pr+qs)-4\varepsilon^2[
p\alpha\eta(qs+r^2+\xi\beta+\alpha\eta+pr)+r(\alpha\eta +p^2+pr)(sq+\beta\xi)+p\beta\xi(qs+r^2)+q^2s^2(p+r)],$$
$$\alpha_{t_2}=-\frac{1}{4} \beta_{xx}+\frac{1}{2}[(p\beta)_x-(q\alpha)_x]+\frac{1}{2}\beta(p^2+q^2)+\varepsilon \beta(qp_x-pq_x+sr_x-rs_x+\alpha\alpha_x+\eta\eta_x+\xi\xi_x-\alpha_x\xi)$$
$$-\varepsilon \alpha_x(pr+qs-\alpha\eta)+\varepsilon^2\beta[qs(qs+2pr+2\alpha\eta)+pr(pr+2\alpha\eta)],~~~~~~~~~~~~~~~~~~~~~~~~~~~$$
$$\beta_{t_2}=\frac{1}{4}\alpha_{xx}+\frac{1}{2}[(q\beta)_x-(p\alpha)_x]-\frac{1}{2}\alpha(p^2+q^2)+\varepsilon \alpha(pq_x-qp_x-sr_x+rs_x+\eta\eta_x+\xi\xi_x-\beta\beta_x-\beta_x\eta)$$
$$-\varepsilon \beta_x(pr+qs+\xi\beta)-\varepsilon^2\alpha[pr(pr+2qs+2\beta\xi)+qs(qs+2\beta\xi)],~~~~~~~~~~~~~~~~~~~~~~~~~~~~~~$$
$$\xi_{t_2}=-(\eta_{xx}+\beta_{xx})+\frac{1}{2}[2p\beta_x+\beta p_x-q\alpha_x-2\alpha q_x+(\beta r)_x-(\alpha s)_x+\eta p_x+2p\eta_x-\xi q
_x-2q\xi_x]~~$$
$$+\frac{1}{2}(\beta+\eta)(p^2+q^2)+\beta(ps+qr)-2\varepsilon\alpha_x(qs+pr-\beta\xi+\frac{1}{2}\alpha\eta)-2\varepsilon\xi_x(qs+pr+\beta\xi$$$$~~~+
\alpha\eta+\frac{1}{2}\eta\xi)+\varepsilon\eta(qp_x-pq_x+sr_x-rs_x+\beta\beta_x)+\varepsilon^2[
\beta pr(pr+2qs+2\xi\eta+2\alpha\eta+pr)$$$$+\beta qs(\alpha\eta+\xi\eta+qs)+\eta(p^2r^2+q^2s^2)+2\eta pqrs],~~~~~~~~~~~~~~~~~~~~~~~~~~~~~~~~~~~~~~~~~~~~$$
$$\eta_{t_2}=\alpha_{xx}+\xi_{xx}+\frac{1}{2}[2p\alpha_x+2q\beta_x+\beta q_x+\alpha p_x+(\beta s)_x+(\alpha r)_x+\eta q_x+2q\eta_x+r\alpha
_x+\xi p_x+2p\xi_x]$$
$$~~~~~~~+\frac{1}{2}(\alpha-\xi)(p^2+q^2)+\alpha(ps+qr+\frac{1}{2}\beta\xi)-2\varepsilon\beta_x(qs+pr-\alpha\eta+\frac{3}{2}\xi\beta)
-2\varepsilon\eta_x(qs+pr+\beta\xi$$$$~~~~~~~~~~~~+
\alpha\eta+\frac{1}{2}\eta\xi)-\varepsilon\xi(qp_x-pq_x-sr_x+rs_x+\alpha\alpha_x)-\varepsilon^2[
\alpha pr(qs+pr+2\beta\xi+2\eta\xi)+\alpha qs(2\beta\xi+2\eta\xi+qs)$$$$+\xi(p^2r^2+q^2s^2)+2\xi pqrs].~~~~~~~~~~~~~~~~~~~~~~~~~~~~~~~~~~~~~~~~~~~~~~~~~~~~~~~~~~~~~~~~~~~~~~~~~$$

\subsection{Completion of super NLS-mKdV integrable couplings}
Similar to the case of the super dirac hierarchy, in this section, we will briefly show how to construct the complete system of the super  NLS-mKdV  integrable couplings. Firstly,  we consider the following super  spectral problem
\begin{equation}
\label{eqns:0nls}
\phi_x=\overline{U}\phi,~\bar{U}=\left[
      \begin{array}{ccc|ccc}
     \lambda +h &   p+q &\alpha &  0&r+s&\xi\\
       -p+q &  -\lambda -h& \beta &-r+s&0&\eta\\
      \beta&-\alpha& 0 &\eta &-\xi&0\\ \hline
      0&0&  0&  \lambda +h &p+q &\alpha\\
             0&0&  0&  -p+q & -\lambda -h& \beta \\
         0&0&  0&\beta &-\alpha & 0
      \end{array}
    \right],~~~h=\varepsilon(qs-pr+\alpha\eta+\xi\beta).
\end{equation}
with $p,~q,~r,~s,~\alpha,~\beta,~\xi,$  and $\eta$ are all functions of $x$ and $t$, $p,~q,~r,~s$ , $\lambda$ are
bosonic, and $\alpha,~\beta,~\xi$, $\eta$ are fermionic.
Here, $h$ is an added  nonlinear perturbation term. When $\varepsilon=0$, it reduces the case of super NLS-mKdV  integrable couplings\cite{smkdv1}.

Upon setting
\begin{equation}
\label{eqns:wnls}
\overline{W}=\left[
      \begin{array}{ccc|ccc}
       a &b+c &\rho &d&f+g& \mu\\
     b-c & -a &\sigma &f-g & -d &\nu \\
     \sigma&-\rho &0 &\nu &-\mu & 0\\ \hline
      0&0&0 & a &b+c &\rho\\
      0&0&0 &b-c & -a &\sigma \\
       0&0&0&\sigma&-\rho &0
      \end{array}
    \right]=\sum_{i\geq 0}\left[
      \begin{array}{ccc|ccc}
       a_i &b_i+c_i &\rho_i &d_i&f_i+g_i& \mu_i\\
     b_i-c_i & -a_i &\sigma_i &f_i-g_i & -d_i &\nu_i \\
     \sigma_i&-\rho_i &0 &\nu_i &-\mu_i & 0\\ \hline
      0&0&0 & a_i &b_i+c_i &\rho_i\\
      0&0&0 &b_i-c_i & -a_i &\sigma_i \\
       0&0&0&\sigma_i&-\rho_i &0
      \end{array}
    \right]\lambda^{-i}.
\end{equation}
The stationary zero curvature equation
$
\overline{W}_x=[\overline{U},\overline{W}]
$ gives rise to
\begin{equation}
\label{eqns:dtnls}
\left\{
       \begin{array}{ll}

       a_{i+1,x}=2pb_{i+1}-2qc_{i+1}+\alpha \sigma_{i+1}+\beta \rho_{i+1},\\
              b_{i+1}=\frac{1}{2}c_{ix}+qa_i+\frac{1}{2}\alpha \rho_i+\frac{1}{2}\beta \sigma_i-hb_i,\\
       c_{i+1}=\frac{1}{2}b_{ix}+pa_i+\frac{1}{2}\alpha \rho_i-\frac{1}{2}\beta \sigma_i-hc_i,\\
             \sigma_{i+1}=-\sigma_{ix}-(p-q)\sigma_i+\alpha(c_i-b_i)+\beta a_i-h\sigma_i,\\
             \rho_{i+1}=\rho_{ix} -(p+q)\rho_i+\alpha a_i+\beta (b_i+c_i)-h\rho_i,\\
             d_{i+1,x}=-2sc_{i+1}+2rb_{i+1}+2pf_{i+1}-2qg_{i+1}+\alpha \nu_{i+1}+\xi \sigma_{i+1}+\xi \sigma_{i+1}+\beta \mu_{i+1},\\
    f_{i+1}= \frac{1}{2}  g_{ix}-hf_i+q d_i+(r+s)a_i -\frac{1}{2} \sigma_i \eta-\nu_i\beta-\mu_i \alpha-\rho_i \xi,\\
     g_{i+1}=\frac{1}{2} f_{ix}-hg_i+pd_i-\frac{1}{2}\eta \sigma_i-\frac{1}{2}\beta \nu_i-\frac{1}{2}\mu_i\alpha-\frac{1}{2}\rho_i\xi,\\
       \mu_{i+1}=\mu_{ix}-(p+q)\nu_i-h\mu_i-(r+s)\sigma_i+\beta (f_i+g_i)+\alpha d_i+\eta (b_i+c_i)+\xi a_i,\\
     \nu_{i+1}=- \nu_{ix}+(p-q)\mu_i-h\mu_i+(s-r)\rho_i+\beta d_i-\alpha (f_i-g_i)+\xi (c_i-b_i)+\eta a_i.
       \end{array}
     \right.
\end{equation}
This system can uniquely determine all sets of functions of $a_i,~b_i,~c_i,~d_i,~f_i,~g_i,~\rho_i,~\sigma_i,~\mu_i$ and $\nu_i$
 upon choosing the initial values as
 $$b_0=c_0=f_0=g_0=\rho_0=\sigma_0=\mu_0=\nu_0=0,~a_0=1,~d_0=1$$
 and assuming $a_i\mid_{u=0}=b_i\mid_{u=0}=c_i\mid_{u=0}=d_i\mid_{u=0}=f_i\mid_{u=0}=g_i\mid_{u=0}\rho_i\mid_{u=0}=\sigma_i\mid_{u=0}
 =\mu_i\mid_{u=0}=\nu_i\mid_{u=0}=0,~i\geq 1$.

  The first few sets are:
  $$a_1=0,~b_1=q,~c_1=p,~\rho_1=\alpha,~\sigma_1=\beta,~d_1=0,~f_1=q+s,~g_1=p+r,~\mu_1=\xi+\alpha,~\nu_1=\eta+\beta,$$
  $$a_2=\frac{1}{2}(p^2-q^2)-\alpha\beta,~b_2=\frac{1}{2}p_x-\varepsilon q(qs-pr+\alpha \eta+\xi \beta),~c_2=\frac{1}{2}q_x-\varepsilon p(qs-pr+\alpha \eta+\xi \beta),~~$$
 $$\rho_2=\alpha_x-\varepsilon \alpha(qs-pr+\alpha \eta+\xi \beta),~\sigma_2=-\beta_x-\varepsilon \beta(qs-pr+\alpha \eta+\xi \beta),~~~~~~~~~~~~~~~~~~~~~~~~~~~~~~~~$$
 $$f_2=\frac{1}{2}(r_x+p_x)-\beta\eta-\alpha\xi-\varepsilon (q+s)(qs-pr+\alpha \eta+\xi \beta),~g_2=\frac{1}{2}(s_x+q_x)+\beta\eta+\alpha\xi-\varepsilon (p+r)(qs-pr+\alpha \eta+\xi \beta),$$
 $$\mu_2=\alpha_x+\xi_x-\varepsilon (qs-pr+\alpha \eta+\xi \beta)(\alpha+\xi),~\nu_2=-\eta_x-\beta_x-\varepsilon (qs-pr+\alpha \eta+\xi \beta)(\eta+\beta),~~~~~~~$$
 $$d_2=pr-qs+\frac{1}{2}(p^2-q^2)+\eta\alpha-(\alpha+\xi) \beta.~~~~~~~~~~~~~~~~~~~~~~~~~~~~~~~~~~~~~~~~~~~~~~~~~~~~~~~~~~~~~~~~~~~~$$

For any integer $m\geq1$, we take
\begin{equation}
\label{eqns:vanls}
V^{[m]}=(\lambda^{m}W)_++\triangle_m=\sum_{i=0}^{m}V_i+\triangle_m,~
\triangle_m=\left[
      \begin{array}{cccccc}
       -\varepsilon d_{m+1} &0 &0&0&0& 0\\
     0 & \varepsilon d_{m+1} &0&0&0&0 \\
     0&0&0& 0&0&0\\
      0&0&0 &  -\varepsilon d_{m+1} &0  &0\\
      0&0&0 &0 & \varepsilon d_{m+1} &0 \\
       0&0&0&0&0 &0
      \end{array}
    \right].
\end{equation}
 Thus, the corresponding zero curvature equation
 (\ref{eqns:zce}) with $\overline{U},~V$ defined in (\ref{eqns:0nls}) and (\ref{eqns:vanls})
determine the complete system of the  super NLS-mKdV  integrable couplings
\begin{equation}
\label{eqns:cs}\left[
      \begin{array}{c}
       p_{tm}\\
    q_{tm} \\
     r_{tm}\\
    s_{tm}\\
    \alpha_{t_m}\\
   \beta_{t_m}\\
   \xi_{t_m}\\
   \eta_{t_m}
      \end{array}
    \right]=\left[
      \begin{array}{c}
       2b_{m+1}-2\varepsilon qd_{m+1} \\
      2c_{m+1}-2\varepsilon pd_{m+1}\\
      2f_{m+1}-2\varepsilon sd_{m+1}\\
         2g_{m+1}-2\varepsilon rd_{m+1} \\
   \rho_{m+1}-\varepsilon \alpha d_{m+1}\\
   -\sigma_{m+1}+\varepsilon \beta d_{m+1}\\
   \mu_{m+1}-\varepsilon \xi d_{m+1}\\
   -\nu_{m+1}+\varepsilon \eta d_{m+1}\\
      \end{array}
    \right],~~~~~m\geq 0.
\end{equation}
The first nonlinear super system is
$$p_{t_2}=\frac{1}{2} q_{xx}+\alpha\alpha_x-\beta\beta_x+q(p^2-q^2-2\alpha \beta)-2\varepsilon[q(\xi\beta)_x+(\eta\alpha)_x+qp_x-pq_x+\alpha\beta_x+\beta\alpha_x~~~~$$$$~~+\frac{1}{2}(sp_x-qr_x+rq_x)-sp_x+qp_x+ps_x)+p_x(-pr+\xi\beta+\eta\alpha)]
+4\varepsilon^2[-q^2s(
pr+\alpha\beta)$$$$+q^2(ps-\eta\beta+\xi\alpha)+q\beta(\xi\eta\alpha-pr\xi+p(\xi-\eta)-p\alpha-p\alpha(q(\xi+\eta)+rq\eta)],~~~~~~~~~~$$
$$q_{t_2}=\frac{1}{2} p_{xx}+\alpha\alpha_x+\beta\beta_x+p(p^2-q^2-2\alpha \beta)-2\varepsilon[p(\eta\alpha_x+\alpha\eta_x-qp_x+pq_x+\beta\xi_x+\xi\beta_x~~~~~~~~~$$$$+\beta\alpha_x+\alpha\beta_x-\frac{1}{2}((qr)_x+sp_x-ps_x)+2q\eta\beta
+2(p-q)\xi\alpha)+q_x(qs
+\xi\beta+\eta\alpha)]~~~~~$$$$
+4\varepsilon^2[p^2(
ps-\alpha\eta+\xi\beta-\alpha\beta-\frac{1}{2}pr^2)pqs(-\alpha\eta-\alpha\beta-\beta\xi)-\frac{1}{2}pq^2s^2],~~~~~~~~~~~~~~~~~~$$
$$\alpha_{t_2}= \alpha_{xx}+\frac{1}{2}(\beta q_x+2q\beta_x+\beta p_x+2p\beta_x)+\frac{1}{2} \alpha(p^2-q^2)+\frac{1}{2}\varepsilon[\alpha(rq_x-r_xq+ps_x-sp_x-2qp_x+2pq_x$$$$+2\beta\alpha_x+2\beta\xi_x+2\xi\beta_x-2\alpha_x\eta)+4\alpha_x(pr-qs-\xi\beta)
+4\alpha\eta\beta(p+q)]~~~~~~~~~~~~~~~~~~~~~$$$$
+2\varepsilon^2\alpha[ pr(
pr-qs-\beta\xi)+qs(qs+2\beta\xi)-p(qs+\beta\xi)],~~~~~~~~~~~~~~~~~~~~~~~~~~~~~~~~~~~~~$$
$$\beta_{t_2}= -\beta_{xx}+\frac{1}{2}\alpha( p_x-q_x)+\alpha_x(p-q) +\frac{1}{2}\beta(q^2-p^2)+\frac{1}{2}\varepsilon[\beta(p_xs-sp_x+qr_x-q_xr-2pq_x-2q_xp-2\beta\alpha_x$$$$-2\eta\alpha_x-2\alpha\eta_x-6\xi\beta_x)
+2\beta_x(pr-qs-\eta\alpha)4\beta\xi\alpha(q-p)]
+2\varepsilon^2\beta[ pr(
qs-pr+\eta\alpha)~~$$$$+qs(\beta p-qs-2\eta\alpha)-\eta\alpha p],~~~~~~~~~~~~~~~~~~~~~~~~~~~~~~~~~~~~~~~~~~~~~~~~~~~~~~~~~~~~~~~~~~~~~~$$
$$r_{t_2}= \frac{1}{2}(q_{xx}+s_{xx})+2(\beta\eta)_x+ \beta\beta_x-\alpha\alpha_x-q(3qs-q^2+p^2+2pr-2\alpha\beta+2\eta\alpha-2\xi\beta)+s(p^2-2\alpha\beta)+\varepsilon[s(q_xr
$$$$~~~~~~~~~~~~-3qr_x-4qp_x)+2s(ps_x+pq_x+\alpha\beta_x+\beta\alpha_x+\eta\alpha_x+\alpha\eta_x+\beta\xi_x+\xi\beta_x-\frac{1}{2}sp_x)+2(pr-\eta\alpha-\xi\beta)(r_x+p_x)
$$$$~~~~~~~~~~~~-2(pr+ps)(\alpha\xi+\beta\eta)+2ps(\eta\beta+\xi\alpha)+4\eta\beta s(p+q)
+4\varepsilon^2[ p^2r^2(s+q)
+2q^2s^2(q+2s)+2s^2q(\alpha\beta+2\beta\xi$$$$~~~~~~~~~~~~+2\alpha\eta-2pr)+4sq^2(\alpha\eta+\beta\xi-pr)-4r(sp+pq)(\alpha\eta+\beta\xi)-4ps(\alpha\eta+\beta\xi+qs)
+4\alpha\beta(ps+\eta\xi(q+s))],$$
$$s_{t_2}= \frac{1}{2}(p_{xx}+r_{xx})-2(\beta\eta)_x- \beta\beta_x+\alpha\alpha_x
+p(3pr-q^2+p^2-2qs-2\xi\beta+2\eta\alpha-2\alpha\beta)-r(q^2
+2\alpha\beta)+\varepsilon[r(-r_xq$$$$~~~~~~~~~~~+rq_x
-p_xs-2qp_x+2\alpha\beta_x+2\beta\alpha_x+2\eta\alpha_x+2\alpha\eta_x+2\beta\xi_x+2\xi\beta_x+2pq_x+3ps_x)-2(qs+\eta\alpha+\xi\beta)(q_x+s_x)
$$$$~~~~~-4q(r+s)(\xi\alpha+\beta\eta)+8pr(\beta\eta+\alpha\xi)]
+4\varepsilon^2[
2(q^2s^2+p^2r^2)(p+2r)-4r^2p(\alpha\eta+\beta\xi+qs)$$$$~~~~~~~~~-4rp^2(\alpha\eta+\beta\xi+qs)+4qs(p+2r)(\alpha\eta+\beta\xi)+4pr(\alpha\beta+\alpha\eta-\beta\xi-qs)
+4\alpha\beta(rqs+\eta\xi(p+r))],$$
$$\mu_{t_2}= \alpha_{xx}+\xi_{xx}+\alpha(pr-qs-2\xi\beta-\frac{1}{2}q^2+\frac{1}{2}p^2)+\frac{1}{2}\xi(p^2-q^2)+(\beta_x+\eta_x)(p+q)+\frac{1}{2}(\eta+\beta)
(p_x+q_x)
$$$$~~~~~~~~+\beta_x(r+s)+\frac{1}{2}\beta(r_x+s_x)+\varepsilon[\xi(\alpha\beta)_x+\eta\alpha_x+\alpha\eta_x-\beta\xi_x+pq_x-qp_x+\frac{1}{2}(q_xr-qr_x+ps_x-p_xs))
$$
$$~~~~~~~~~+(\alpha_x+\xi_x)(pr-qs)-2\eta\alpha(\alpha_x+\xi_x)+2\xi\eta\beta(p+q)]
-\varepsilon^2[2(q^2s^2+p^2r^2)(\xi+\alpha)+2\alpha pr(qs+\xi\eta-\beta\xi)$$$$+2\alpha\xi (qs+p)(\eta+\beta)-\xi pqs(r+1)],~~~~~~~~~~~~~~~~~~~~~~~~~~~~~~~~~~~~~~~~~~~~~~~~~~~~~~~~~~~~~~~~$$
$$\nu_{t_2}= -\beta_{xx}-\eta_{xx}+\frac{1}{2}(\beta+\eta)(q^2-p^2)+\frac{1}{2}(\alpha+\xi)(p_x-q^2)+(\alpha_x+\xi_x)(p-q)+\alpha_x(r-s)
+\frac{1}{2}\beta(r_x-s_x)$$$$~~~~~~~~~-\beta(pr-qs+2\alpha\eta)+\varepsilon[-\eta(\beta\alpha_x+3\alpha\beta_x+3\alpha\eta_x+\beta\xi_x+\xi\beta_x+(pq)_x+
\frac{1}{2}(qr_x-q_xr+p_xs-ps_x))
$$
$$-2(\beta_x+\eta_x)(qs+\beta\xi-pr)-3]
+\varepsilon^2[-(q^2s^2+p^2r^2)(\beta+2\eta)+2\eta pr(qs+\xi\beta+\alpha\beta)~~$$$$-4\eta qs (\xi\beta+\alpha\beta)+2\eta p(qs+\xi\beta-\alpha\beta+\alpha\xi)+\eta\xi\alpha q].~~~~~~~~~~~~~~~~~~~~~~~~~~~~~~~~~~~~~~~~$$

\section{Conclusion}
 To present, there many methods to generate integrable couplings. In generally, the integrable couplings is a triangular integrable system:
  $$
\left\{
       \begin{array}{ll}
       u_t=K(u), \\
       v_t=S(u,v).
       \end{array}
     \right.
$$
In this system, the first equation is about $u$ and the second equation is about $u$ and $v$. At present paper, we construct a kind of new coupled system
\begin{equation}
\label{eqns:gc}
\left\{
       \begin{array}{ll}
       u_t=\overline{K}(u,v), \\
       v_t=\overline{S}(u,v),
       \end{array}
     \right.
\end{equation}
by adding a perturbation term $h$ in the spectral matrix, where $h$ may be the combination of potential function and its derivatives.
We call the system   (\ref{eqns:gc})   "completion of the integrable couplings ". For this system, its Hamiltonian structure can also be constructed by
variable identity.  This method  can be applied to  continuous integrable couplings, discrete integrable couplings and super integrable couplings.
As an example, we generate the  complete integrable system of the KN integrable coupling and construct  its Hamiltonian structure by variable identity.
As application, we also construct
 the completion of the
KN integrable coupling,   the WKI integrable couplingsis, vAKNS integrable couplings, the Volterra integrable couplings, Dirac type integrable couplings and NLS-mKdV type integrable couplings.

\section*{Acknowledgement}
This work is in part supported by the national natural science foundation of China (Grant No 11371323, 11371326 and 11271226) and Beijing Municipal Natural Science Foundation (Grand No 1162003).

\end{document}